\documentclass[a4paper,onecolumn, 11pt, accepted=2022-07-06]{quantumarticle}
\pdfoutput=1

\usepackage{amsfonts,amsthm,xspace,graphicx,relsize,bm,mathtools,xcolor,amsmath,float}
\usepackage{bbm}
\usepackage{mathrsfs}

\newtheorem{theorem}{Theorem}[section] 

\newtheorem{lemma}[theorem]{Lemma}

\newtheorem{fact}[theorem]{Fact}
\newtheorem{corollary}[theorem]{Corollary}
\newtheorem{definition}[theorem]{Definition}
\newcommand{\E}{\mathbb{E}}

\def\01{\{0,1\}}
\newcommand{\ceil}[1]{\lceil{#1}\rceil}
\newcommand{\floor}[1]{\lfloor{#1}\rfloor}

\newcommand{\ket}[1]{|#1\rangle}

\newcommand{\ketbra}[2]{|#1\rangle\langle#2|}
\newcommand{\braket}[2]{\langle#1|#2\rangle}

\renewcommand{\>}{\rangle}
\newcommand{\<}{\langle}

\newcommand{\Exp}{\mathbb{E}}

\newcommand{\varS}{\ensuremath{\mathbb{S}}}

\newcommand{\Zp}{\ensuremath{\mathsf{Z}}}

\newcommand{\A}{\ensuremath{\mathcal{A}}}

\newcommand{\id}{\ensuremath{\mathbb{I}}}

\def\01{\{0,1\}}

\usepackage[noline, boxed]{algorithm2e}

\begin{document}

\title{Simpler (Classical) and Faster (Quantum) Algorithms for Gibbs Partition Functions}

\author{Srinivasan Arunachalam}
\affiliation{IBM Quantum, 
IBM T.J. Watson Research Center}

\author{Vojtech Havlicek}
\affiliation{IBM Quantum, 
IBM T.J. Watson Research Center}
\affiliation{School of Mathematics, University of Bristol}

\author{Giacomo Nannicini}
\affiliation{IBM Quantum, 
IBM T.J. Watson Research Center}

\author{Kristan Temme}  
\affiliation{IBM Quantum, 
IBM T.J. Watson Research Center}

\author{Pawel Wocjan}
\affiliation{IBM Quantum, 
IBM T.J. Watson Research Center}

\maketitle

\begin{abstract}
We present classical and quantum algorithms for approximating partition functions of classical Hamiltonians at a given temperature. Our work has two main contributions: first, we modify the classical algorithm of \v{S}tefankovi\v{c}, Vempala and Vigoda (\emph{J.~ACM}, 56(3), 2009) to improve its sample complexity; second, we quantize this new algorithm, improving upon the previously fastest quantum algorithm for this problem, due to Harrow and Wei (SODA 2020). The conventional approach to estimating partition functions requires approximating the means of Gibbs distributions at a set of inverse temperatures that form the so-called cooling schedule. The length of the cooling schedule directly affects the complexity of the algorithm. Combining our improved version of the algorithm of \v{S}tefankovi\v{c}, Vempala and Vigoda with the paired-product estimator of Huber (\emph{Ann.\ Appl.\ Probab.}, 25(2),~2015), our new quantum algorithm uses a shorter cooling schedule than previously known. This length matches the optimal length conjectured by  \v{S}tefankovi\v{c}, Vempala and Vigoda. The quantum algorithm also achieves a quadratic advantage in the number of required quantum samples compared to the number of random samples drawn by the best classical algorithm, and its computational complexity has quadratically better dependence on the spectral gap of the Markov chains used to produce the quantum samples.

\end{abstract}

\noindent
\emph{Note: a shorter version of this work previously appeared in the proceedings of the 2021 IEEE International Conference on Quantum Computing and Engineering (QCE).}

\medskip
\newpage 

\tableofcontents

\section{Introduction}
Markov Chain Monte Carlo (MCMC) method is a fundamental computational tool in statistics. It is a strategy for sampling from certain high-dimensional probability distributions, which can be used to estimate properties of systems that would be difficult to study otherwise. In the last two decades, MCMC methods found numerous applications in Bayesian inference~\cite{goodfellow16,gelman1998simulating}, counting problems~\cite{vstefankovivc2009adaptive}, volume estimation of convex bodies~\cite{dyer1991volume}, approximation of the permanent~\cite{jerrum1989approximating}, estimating thermodynamic properties of systems~\cite{binder19} or in finance for simulating the performance and volatility of  portfolios~\cite{glasserman2013monte}.

A major algorithmic task in the area of MCMC methods is the approximation of~\emph{partition functions}. In physics, the partition function describes the statistical properties of a physical system at a fixed inverse temperature, but a wide range of problems can be naturally cast as questions about partition functions. In machine learning for example, the partition function appears in the definition of probabilistic graphical models such as Markov Random Fields (e.g., restricted Boltzman machines \cite{hinton2006fast}), and has been extensively studied before~\cite{desjardins2011tracking,liu2011bounding,huber2015approximation,Kolmogorov18}. Given the many of their applications, it is important to design efficient algorithms for their computation, and explore the potential of using faster techniques to compute them.

We study the advantages that quantum computers have over classical computers for the computation of partition functions.  
It is well known that quantum algorithms provide quadratic speedups over classical algorithms for a variety of tasks related to MCMC, such as amplitude amplification and estimation~\cite{brassard2002quantum} and spectral gap amplification of Markov Chains~\cite{szegedy2004quantum}. Even though a general quantum speedup for MCMC methods has not been found,  fast quantum algorithms for partition functions using simulated annealing have been proposed~\cite{wocjan2008speedup,montanaro2015quantum,harrow2020adaptive}. Here we study classical and quantum algorithms for computing partition functions; our main contribution is a unification of two \emph{classical} algorithms, leading to a \emph{quantum} algorithm that improves upon the best known (quantum or classical)~approaches. Before we describe our main results, we first introduce some basic notation and the problem of computing Gibbs partition functions.

\subsection{Estimating partition functions}
Let $\Omega$ be a finite set and
$H:\Omega\rightarrow \{0,\dots,n\}$ be a function called the
\emph{Hamiltonian}\footnote{Hamiltonians are usually defined as real-valued functions. In this paper we make the simplifying assumption that the range of the Hamiltonian is contained in the set of nonnegative integers. This assumption is common in the literature and can often be relaxed, see Kolmogorov~\cite{Kolmogorov18} for a discussion.}. The \emph{Gibbs distribution} is defined as:
\begin{align*}
    \mu_\beta(x) &= \frac{1}{\Zp(\beta)} \exp(-\beta H(x)) \quad \text{ for every } x\in \Omega,
\end{align*}
where $\beta$ is the
\emph{inverse temperature} and 
$
\Zp(\beta) := \sum_{x \in \Omega} e^{-\beta H(x)}
$
is the \emph{partition function}.  Given $\beta_{\max}$ and $\varepsilon > 0$, we aim to find $\hat{Z}$ such that
$
   (1-\varepsilon) \cdot \Zp(\beta_{\max}) \leq \hat{Z} \leq (1+\varepsilon) \cdot \Zp(\beta_{\max})
$
using the least amount of samples from $\mu_\beta$ for various choices of $\beta$. 
An approach for this task is based on the following telescoping product:
 \begin{equation}
\begin{aligned}
\label{eq:introtelescoping}
\Zp(\beta_{\max})&=\Zp(0)\cdot \frac{\Zp(\beta_1)}{\Zp(0)}\cdot \frac{\Zp(\beta_2)}{\Zp(\beta_1)}\cdots \frac{\Zp(\beta_{\max})}{\Zp(\beta_{\ell-1})} =\Zp(0)\cdot \prod_{i=0}^{\ell-1} \frac{\Zp(\beta_{i+1})}{\Zp(\beta_{i})},
\end{aligned}
\end{equation}
where $0=\beta_0 < \beta_1 < \dots < \beta_{\ell} = \beta_{\max}$ 
is a sequence of inverse
temperatures called a \emph{cooling schedule}.
 Eq.~\ref{eq:introtelescoping} approximates $\Zp(\beta_{\max})$ if each factor $\Zp(\beta_{i+1})/\Zp(\beta_{i})$ is estimated with sufficient precision. We fix $\beta_{\max} = \infty$ in the rest of the introduction for simplicity. 

Two approaches for this approximation were introduced in Refs.~\cite{vstefankovivc2009adaptive,huber2015approximation}. The algorithm of {\v{S}}tefankovi{\v{c}} et al.~\cite{vstefankovivc2009adaptive} computes $\Zp(\infty)$ in two steps: first, it produces a cooling
schedule consisting of $O(\sqrt{\ln |\Omega|} \ln n \ln \ln |\Omega|)$ inverse temperatures. Defining $X_i=e^{(\beta_i-\beta_{i+1})H(x)}$, we have
$\E_{x\sim \mu_{\beta_i}}  [X_i]= \Zp(\beta_{i+1}) / \Zp(\beta_{i})$.
The partition function can be approximated by sampling many $x\sim \mu_{\beta_i}$, empirically estimating the factors at subsequent temperatures and taking their product. 
This is called the \emph{product estimator}. 
The resulting algorithm for partition function estimation has sample complexity $10^{10}\cdot \ln |\Omega| \cdot \varepsilon^{-2}\cdot \big(\ln n+\ln \ln |\Omega|\big)^5$. We call this algorithm SVV, for \v{S}tefankovi\v{c}, Vempala and Vigoda.

SVV contains many subtle technical points that lead to some of the polylogarithmic factors and a large prefactor in its running time. Huber~\cite{huber2015approximation} and Kolmogorov~\cite{Kolmogorov18} eliminated these hurdles using the so-called Tootsie Pop Algorithm (TPA)~\cite{huber2010using}. This algorithm uses a Poisson point process to build a  cooling schedule with $\sim \ln |\Omega|$ temperatures. The product estimator would be inefficient in this context as it would pick up a quadratic overhead in the schedule length. To address this, Huber introduced the \emph{paired-product estimator}. Given a cooling schedule, define $d_{i,i+1}=(\beta_{i+1}-\beta_{i})/2$, and random variables~$V_i, W_i$:
\begin{equation*}
V_i = \exp\left(-d_{i,i+1} \, H(x_i)     \right),\quad     W_i = \exp\left( d_{i,i+1} \, H(x_{i+1}) \right),
\end{equation*}
where $x_i \sim \mu_{\beta_i}$.
The telescoping product in Eq.~\eqref{eq:introtelescoping} can be rewritten as
\begin{equation}
  \label{eq:introtelehuber}
\Zp(\infty)=\Zp(0)\cdot \prod_{i=0}^{\ell-1} \frac{\Exp[W_i]}{\Exp[V_i]} = \Zp(0)\cdot \frac{\prod_i\Exp[W_i]}{\prod_i\Exp[V_i]}.
\end{equation}
While in SVV each ratio
$\Exp[W_i]/\Exp[V_i]$ is estimated separately, Huber estimates the
entire numerator $\prod_i\Exp[W_i]$ and denominator
$\prod_i\Exp[V_i]$, then takes their ratio. This, remarkably, leads to a better estimator: the overall complexity of
TPA (using the improved analysis given by
Kolmogorov~\cite{Kolmogorov18}) is
$O(\ln |\Omega| \cdot \varepsilon^{-2}\cdot \ln n )$.

\subsection{Results}\label{sec:results}
We give a quantum algorithm that
estimates partition functions by combining elements of SVV and TPA. Assuming the ability to query an oracle that provides coherent encodings of the Gibbs distribution at arbitrary inverse temperature (\emph{qsamples}), we show that the quantum algorithm requires fewer qsamples compared to the number of samples drawn by the best classical algorithm.

The above assumption is not restrictive, as qsamples can be generated from a classical Markov chain that has the corresponding Gibbs distribution as its limit distribution; the total running time of the algorithm (classical and quantum) must also account for the time -- known as \emph{mixing time} -- taken to generate these samples. Quantum algorithms for approximating partition functions~\cite{wocjan2008speedup,montanaro2015quantum,harrow2020adaptive} do not perform the classical steps of the MCMC algorithm. Instead, they utilize unitary quantum walks~\cite{szegedy2004quantum} that are related to the classical Markov chain by a suitable embedding as the product of two reflection operators. Such quantum walks are constructed so that a natural quantum encoding of the Gibbs distribution is an eigenvector with eigenvalue one of the quantum walk unitary. This leads to a quadratic advantage of the quantum algorithm over classical, in terms of the mixing time.

\subsubsection{Classical contribution}
We propose a classical algorithm for the approximation of partition functions that simplifies
\cite{vstefankovivc2009adaptive}, and almost matches the sample complexity of TPA, the best classical algorithm \cite{Kolmogorov18, huber2015approximation}. It uses the paired-product estimator within SVV: this allows for a shorter cooling schedule and a simpler estimator. 

This estimator becomes simpler (motivating the ``simpler (classsical)" in the paper title), because the approximation of the partition function ratios in the cooling schedule algorithm does not depend on temperatures outside the interval on which these are estimated. This gets rid of the nested search in the SVV cooling schedule algorithm, a siginficant simplification. Moreover, unlike TPA, where these ratios were estimated using a Poisson Point Process, our algorithm is easier to quantize.

Ref.~\cite{vstefankovivc2009adaptive} conjectured that an optimal adaptive cooling schedule has length $\Theta(\sqrt{\ln |\Omega| \ln n})$ and proved a lower bound of $\Omega(\sqrt{\ln |\Omega|})$.\footnote{The lower bound   of \cite{vstefankovivc2009adaptive} is for the so-called $B$-Chebyshev schedules, however their proof applies to the \emph{slowly-varying} schedules proposed in this paper, with straightforward modifications leading to a constant-factor~difference.} They give an algorithm that produces a schedule of length $O(\sqrt{\ln |\Omega|} \ln n \ln \ln |\Omega|)$. Our algorithm, albeit quantum, produces a schedule of length $O(\sqrt{\ln |\Omega|} \ln n)$, while achieving the same approximation precision.

\subsubsection{Quantum contribution}
Our main contribution is a quantum algorithm that improves the best known sample complexity and overall running time. In the quantum setting, the samples are the \emph{coherent encodings} of the Gibbs distribution (qsamples):
$$
\ket{\mu_{\beta}}=\sum_{x}\sqrt{\mu_{\beta}(x)}\ket{x}.
$$

The quantum algorithm is based on SVV, but uses the paired-product estimator of \cite{huber2015approximation}. We give a procedure,
based on binary search, to obtain a cooling sequence of length
$O(\sqrt{\ln |\Omega| \ln n})$, using the fact that we can efficiently compute inner products of qsamples of the form
$|\braket{\mu_{\beta_i}}{\mu_{\beta_{i+1}}}|$. The length of this cooling sequence matches the optimal length conjectured in \cite{vstefankovivc2009adaptive}.

\subsubsection{A technical tool} To estimate the telescopic product \eqref{eq:introtelehuber}, we give a quantum algorithm for the estimation of the expected value of random variables with bounded relative variance. Computing expectations of random variables is a fundamental task in statistics, applied probability, and related disciplines. Our contribution in this context is a simpler and faster quantum algorithm for computing the expectation of a random variable, given quantum~samples. 
Recall that, in order to compute the Gibbs partition function, one needs to estimate ratios in the telescopic product \eqref{eq:introtelehuber}. To do this, we give a simple quantum algorithm to estimate the expected value of random
variables with bounded relative variance. In particular, for a distribution $D$, and given a random variable $V\sim D$ satisfying $\Exp[V^2]/\Exp[V]^2\leq B$, we describe a
quantum algorithm that uses $O(B)$ copies of $\ket{\psi_D}=\sum_x \sqrt{D(x)}\ket{x}$, and $\tilde{O}\big(\sqrt{B}/\varepsilon)$ reflections about $\ket{\psi_D}$, to obtain with high probability an $\varepsilon$-relative approximation of $\Exp[V]$; additionally, the algorithm restores one copy of $\ket{\psi}$. The proposed algorithm is based on the work of Montanaro~\cite{montanaro2015quantum} and improves upon its scaling from $O(B/\varepsilon)$ to $O(\sqrt{B}/\varepsilon)$. Our algorithm has essentially the same scaling as the algorithm of Hamoudi and Magniez~\cite{hamoudi2019chebyshev}, but the analysis of our algorithm is simpler.

\subsubsection{Applications} 
 
     \label{sec:application}
        Many computational problems can be encoded into evaluations of the partition function and 
 we briefly sketch out two applications of our algorithm. Our work also applies to other problems, such as counting independent sets, matchings and Bayesian inference as discussed in~\cite{harrow2020adaptive, montanaro2015quantum, vstefankovivc2009adaptive}; we refer the interested reader to these works. As discussed earlier, our classical and quantum improvements over~\cite{harrow2020adaptive, vstefankovivc2009adaptive} directly imply better algorithms for estimating partition functions of these Hamiltonians.

\paragraph{Ferromagnetic Ising Model}
The ferromagnetic Ising model on a graph $G = (V,E)$ can be defined as\footnote{The usual convention is to define the model as
$    H(x) = -\sum_{(i,j) \in E} x_i x_j$ for $x \in \lbrace \pm 1 \rbrace^{|V|}$. Here we use a different convention to ensure that the Hamiltonian is nonnegative, which is an assumption of our classical algorithm and other similar algorithms in the literature (e.g., \cite[Lemma~4.2]{vstefankovivc2009adaptive} does not hold for Hamiltonians that change sign).}
\begin{align*}
    H(x) &:= \sum_{(i,j) \in E} \mathbf{1}_{[x_i \neq x_j]},\; x \in \lbrace 0,1 \rbrace^{|V|} .
\end{align*}

The associated Gibbs distribution can be sampled using a Markov Chain called Glauber dynamics. Convergence of Glauber dynamics is well studied and several criteria for rapid mixing have been found. Montanaro~\cite{montanaro2015quantum}  quotes a result of Mossel and Sly~\cite{mossel13} that proves that the chain mixes in time $O(|V| \ln |V|)$ on general graphs with finite degree for a sufficiently low value of $\beta$. As the mixing time upper-bounds the inverse spectral gap (see \cite[Eq.~6]{mossel13}), it follows that for $|V| = n$, our classical algorithm achieves complexity $\tilde{O}(n^2 \cdot \varepsilon^{-2})$. While there may exist algorithms that perform significantly better, the complexity matches (up to polylogarithmic factors) the best classical upper bound that we are aware of \cite{Kolmogorov18}. 
The quantum algorithm in comparison achieves complexity of $\tilde{O}(n^{3/2} \cdot \varepsilon^{-1})$. Apart from being simpler than the algorithm derived in~\cite{harrow2020adaptive}, our approach improves the complexity by polylogarithmic factors. 

\paragraph{$k$-Colorings}\footnotetext{We remark that complexity of~\cite{bezakova2008accelerating} holds for constant $\varepsilon$.}

The $k$-state Potts model on a graph $G = (V,E)$ has the Hamiltonian 
\begin{align*}
\label{eq:potts}
    H(x) = \sum_{(i,j) \in E} \mathbf{1}_{[x_i = x_j]}, \quad x \in \Omega := \{1,\ldots,k\}^{|V|},
\end{align*}
with the corresponding partition function
\begin{align*}
    \Zp(\beta) &= \sum_{x \in \Omega} \exp(-\beta H(x)) = \sum_{x \in \Omega} \prod_{(i,j) \in E} \exp\left(-\beta \mathbf{1}_{[x_i = x_j]}\right).
\end{align*}
Observe that $\lim_{\beta \rightarrow \infty} \Zp(\beta) := \Zp(\infty) = |\lbrace x : x_i \neq x_j,\; \forall (i,j) \in E \rbrace|$. 
A $k$-coloring of a graph is a map $\phi: V \rightarrow \lbrace 1,2, \ldots k \rbrace$, such that $\phi(i) \neq \phi(j)$ for all $(i,j) \in E$. 
The partition function at $\beta = \infty$ therefore gives the number of proper $k$-colorings of $G$. Vigoda~\cite{vigoda00} gave a Markov chain with mixinng time  $O(nk\log{n})$ whenever $k > 11/6 \cdot d$, where $d$ is the maximum degree of $G$. Our quantum algorithm has $\tilde{O}(n^{3/2} \cdot \varepsilon^{-1})$ complexity for this problem.

\subsubsection{Comparison to prior works}
Table \ref{tab:comparison} summarizes the sample complexity of the algorithms presented in this paper and compares them to the existing results in literature.

{\renewcommand{\arraystretch}{1.4} 
\begin{table}[h]
\centering
\begin{tabular}{|c|c|c|}
\hline
               &Algorithm  & Sample Complexity \\
               \hline
    Bezáková et al.~\cite{bezakova2008accelerating} & Classical &$O(\ln^2 |\Omega| \cdot (\ln n)^2)$\;\footnotemark \\
    SVV~\cite{vstefankovivc2009adaptive} & Classical & $O(\ln |\Omega|(\ln \ln |\Omega| + \ln n)^5 \varepsilon^{-2})$ \\
    Huber~\cite{huber2015approximation} &   Classical &  $O(\ln |\Omega| \ln n \cdot \big(\ln \ln |\Omega| + \ln \ln n + \varepsilon^{-2}\big))$ \\
    Kolmogorov~\cite{Kolmogorov18} &Classical& $O(\ln |\Omega| \ln n \cdot \varepsilon^{-2})$ \\
    Montanaro~\cite{montanaro2015quantum} & Hybrid & {$O(\ln |\Omega|\cdot (\ln \ln |\Omega| + \ln n)^{5/2} \cdot  \varepsilon^{-1})$}\\
    Harrow and Wei~\cite{harrow2020adaptive} &Quantum & {$O(\ln |\Omega|\cdot(\ln \ln |\Omega| + \ln n)^{5/2} \cdot  \varepsilon^{-1})$} \\\hline
    This work &Classical & $O(\ln |\Omega|\ln^2 n \cdot \varepsilon^{-2})$ \\
    This work &Quantum & { $O(\ln |\Omega|\ln n \cdot \varepsilon^{-1})$} \\
    \hline

\end{tabular}
\caption{Comparison of the sample complexity of several algorithms to estimate the partition function. Montanaro's algorithm~\cite{montanaro2015quantum} uses classical SVV to generate a cooling schedule, and a quantum routine to estimate the ratio given a cooling schedule; here we report only the sample complexity of the quantum routine, for the total complexity one should add the complexity of classical SVV schedule generation.
The stated complexity of our classical algorithm holds for $\varepsilon < (\ln |\Omega| + \ln n)^{-1}$; the stated complexity of our quantum algorithm holds for $\varepsilon < \left(\frac{\ln |\Omega| + \ln \ln n}{\ln |\Omega| + \ln n}\right)^2$. For a more precise statement of the complexities we refer the reader to Theorems \ref{thm:classicalmainthm} and \ref{thm:quantummainthm}.
}
\label{tab:comparison}
\end{table}}

\section{Classical algorithm}
\label{Sec:Classical}

We first introduce our notation and describe the improved version of SVV, that is the basis for the quantum algorithm.

Let $H:\Omega\rightarrow \{0,\dots,n\}$ be a classical Hamiltonian. For fixed $\beta_{\min}, \beta_{\max}$, we let $Q = \Zp(\beta_{\max})/\Zp(\beta_{\min})$. 
We often use the shorthand $q = \ln |\Omega|$ and assume that $\ln n \ge 1, \ln q \ge 1, |\Omega| \ge \ln n$ as in \cite{vstefankovivc2009adaptive}.\footnote{This assumption is justified because if $|\Omega|$ is small,~$Q$ could be estimated using a schedule of length $1$} 
We assume that the partition function is non-zero so that the Hamiltonian has at least one state in its ground state (i.e. state with a zero energy). This implies that $|\Omega| \geq \Zp(\beta) = \sum_{x \in \Omega} e^{-\beta H(x)} \geq 1$ for all $\beta \in [0, \infty)$. Often in this paper we assume that  $\beta_{\max}\leq q$; this is not restrictive because after a cooling schedule reaches the inverse temperature $q = \ln |\Omega|$, any larger inverse temperature can be reached in a single step.\footnote{Indeed,  we have
  \begin{equation*}
  \begin{aligned}
      \Zp(\ln |\Omega|) &= \sum_{\substack{x \in \Omega :\\ H(x) = 0}} e^{0} + \sum_{\substack{x \in \Omega :\\ H(x) > 0}} e^{-H(x)\cdot \ln |\Omega| } 
       \le \Zp(\infty) + 1. 
      \end{aligned}
  \end{equation*}}

\subsection{Product Estimator}
We aim to estimate $Q = \Zp(\beta_{\max})/\Zp(\beta_{\min})$ given access to samples from the Gibbs distribution at arbitrary inverse temperatures. A naive estimator of $Q$ can be constructed as~follows:
\begin{itemize}
\item sample  $x\sim \mu_{\beta_{\text{min}}}$ from the Gibbs distribution at $\beta_{\min}$. 
\item compute $X = \exp[(\beta_{\text{min}}-\beta_{\text{max}}) H(x)]$.
\item the expectation of $X$ is $\mathbb{E}[X] = \frac{\Zp(\beta_\text{max})}{\Zp(\beta_\text{min})}$. 
\end{itemize}
Define $\varS[X]=\mathbb{E}[X^2]/\mathbb{E}[X]^2$. With a slight overload of terminology, will call $\varS[X]$ the relative variance.\footnote{It is related to the usual definition of relative variance by a constant. } It is equal to:
\begin{equation}\label{eq:relVarBigJump}
    \mathbb{S}[X]=\frac{ \Zp(2 \beta_{\max} - \beta_{\min}) \Zp(\beta_{\min})}{\Zp(\beta_{\max})^2}. 
\end{equation}

This is usually prohibitively large. To address this,
Valleau and Card introduced the \textit{product estimator} in Ref.~\cite{valleau1972monte} as a low-relative variance estimator of $Q$ . It uses a \textit{cooling schedule} of inverse temperatures $\beta_{\min} = \beta_0 < \beta_1 < \cdots < \beta_\ell = \beta_\text{max}$. For every step $0 \leq i < \ell$: 
\begin{itemize}
\item Sample $x \sim \mu_{\beta_{i}}$.
\item Compute $X_i = \exp[(\beta_{i}-\beta_{i+1}) H(x)]$.
\item The expectation of $X$ is $\mathbb{E}[X_i] = \Zp(\beta_{i+1}) / \Zp(\beta_i)$.
\end{itemize}
Let $X$ be the product random variable defined as $X=\prod_{i=1}^{\ell-1} X_i$. The random variables $X_i, X_j$ are independent for $i \neq j$ and it follows that:
\begin{equation*}
\begin{aligned}
    \mathbb{E}[X] &= \prod_{i=0}^{\ell-1} \mathbb{E}[X_i] = \prod_{i=0}^{\ell-1} \frac{\Zp(\beta_{i+1})}{\Zp(\beta_i)} = \frac{\Zp(\beta_{\text{max}})}{\Zp(\beta_{\text{min}})}. 
    \end{aligned}
\end{equation*}
The relative variance of $X_i$ for each $i\in\{0,\ldots,\ell-1\}$ is:
\begin{equation}\label{eq:relVarSmallJumps}
    \mathbb{S}[X_i] = \frac{\mathbb{E}[X_i^2]}{\mathbb{E}[X_i]^2} = \frac{\Zp(2 \beta_{i+1} - \beta_i) \Zp(\beta_i)}{\Zp(\beta_{i+1})^2},
\end{equation}

The product estimator reduces 
the relative variance $\mathbb{S}[X]$ in~\eqref{eq:relVarBigJump}. However, 
the dependence of $\mathbb{S}[X_i]$ on the temperature $2 \beta_{i+1} - \beta_i$, which is outside the interval $[\beta_i,\beta_{i+1}]$, leads to several complications in the analysis of SVV~\cite{vstefankovivc2009adaptive}. 

\subsection{Paired-product Estimator}
\label{sec:pairproduct}
Our algorithm replaces the \emph{product estimator} with the \emph{paired-product estimator}. This estimator was introduced by Huber in~\cite{huber2015approximation}. For $i\in\{0,\ldots,\ell-1\}$, define the midpoint temperature $\bar{\beta}_{i,i+1}=\frac{\beta_i+\beta_{i+1}}{2}$ and 
the semi-distance $d_{i,i+1}=\frac{\beta_{i+1}-\beta_i}{2}$. Then:

\begin{itemize}
    \item For $i\in\{0,\ldots,\ell\}$, sample $x_i \sim \mu_{\beta_i}$.
    \item For $i\in\{0,\ldots,\ell-1\}$, compute
    \begin{equation}
    \begin{aligned}
        V_i &= \exp\big(-d_{i,i+1} \, H(x_i)     \big), &
        W_i &= \exp\big( d_{i,i+1} \, H(x_{i+1}) \big);
        \end{aligned}
    \end{equation}
    observe that $x_i$ is used for $V_i$ and $x_{i+1}$ for $W_i$.
    \item The expectations of $V_i$ and $W_i$ are:
    \begin{align}
        \mathbb{E}[V_i] &= \frac{\Zp(\bar{\beta}_{i,i+1})}{\Zp(\beta_i)}, & 
        \mathbb{E}[W_i] &= \frac{\Zp(\bar{\beta}_{i,i+1})}{\Zp(\beta_{i+1})};
    \end{align}
    observe that $\frac{\mathbb{E}[V_i]}{\mathbb{E}[W_i]} = \frac{\Zp(\beta_{i+1})}{\Zp(\beta_i)}$, i.e., the pair $V_i$ and $W_i$ replaces $X_i$.
\end{itemize}
Define $V_i=\prod_{i=0}^{\ell-1} V_i$ and $W=\prod_{i=0}^{\ell-1} W_i$.  Since the $V_i$ are independent and so are the $W_i$, the ratio of $\mathbb{E}[V]$ and $\mathbb{E}[W]$ can be expressed as:
\begin{equation}\label{eq:ratioEV_EW}
    \frac{\mathbb{E}[V]}{\mathbb{E}[W]} =
    \prod_{i=0}^{\ell-1} \frac{\mathbb{E}[V_i]}{\mathbb{E}[W_i]} =
    \prod_{i=0}^{\ell-1} \frac{\Zp(\beta_{i+1})}{\Zp(\beta_i)} = 
    \frac{\Zp(\beta_{\max})}{\Zp(\beta_{\min)}}.
\end{equation}
The advantage of using the pairs $V_i$ and $W_i$ becomes evident from their relative variances: 
\begin{equation}
\label{eq:relvar}
    \mathbb{S}[V_i] = \frac{\mathbb{E}[V_i^2]}{\mathbb{E}[V_i]^2} = 
    \frac{\Zp(\beta_i) \Zp(\beta_{i+1})}{\Zp(\bar{\beta}_{i,i+1})^2} = 
    \frac{\mathbb{E}[W_i^2]}{\mathbb{E}[W_i]^2} = \mathbb{S}[W_i],
\end{equation}
Both $\mathbb{S}[V_i]$ and $\mathbb{S}[W_i]$ depend on the midpoint $\bar{\beta}_{i,i+1}\in[\beta_i,\beta_{i+1}]$ and it holds that $\mathbb{S}[W_i]=\mathbb{S}[V_i] < \mathbb{S}[X_i]$.\footnote{This follows from the fact that the relative variance is a monotone increasing function in $\beta$. A ``quantum'' proof of this fact is given in Fact~\ref{fact:variances_overlap}.} 

\subsection{Dyer and Frieze's bound on product estimation}
A key technical ingredient in our classical algorithm is a result od Dyer and Frieze \cite{dyer1991random} that bounds the number of samples to estimate the expectation of a product random variable with relative error. 
We use this to estimate the expectation of the product of the random variables $V_i$ and $W_i$. 
\begin{theorem}[Dyer and Frieze \cite{dyer1991random}]
\label{thm:dyerfrieze}
Let $B>0$, $\eta \in (0,1)$. Assume that the independent random variables $X_1,\ldots,X_\ell$ satisfy $\varS [X_i] \le B$ for all $i\in[\ell]$. By taking $m = 2B \ell/(\eta \varepsilon^{2})$    
samples from $X_i$ for every $i\in [\ell]$, we can obtain $\widehat{X}$ that satisfies
\begin{equation*}
\Pr\left[
(1-\varepsilon)\cdot \prod_i\E[X_i] \le \widehat{X} \le (1+\varepsilon)\cdot \prod_i\E[X_i]
\right]
\ge 1-\eta.
\end{equation*}
\end{theorem}

\subsection{Cooling schedule}
We give a short cooling schedule such that the resulting relative variances of the random variables $V_i$ and $W_i$ are bounded from above by a constant. This allows us to apply Theorem~\ref{thm:dyerfrieze}.
To do this, we use a result due to \v{S}tefankovi\v{c} et al.~\cite[Lemma 4.3]{vstefankovivc2009adaptive}, establishing the existence of a short cooling schedule that satisfies: 
\begin{equation}\label{eq:relvar_e2}
    \frac{\Zp(\beta_i) \Zp(\beta_{i+1})}{\Zp(\bar{\beta}_{i,i+1})^2} \le e^2,
\end{equation}
for all $i\in\{0,\ldots,\ell-1\}$. This, together with Eq.~\eqref{eq:relvar}, implies that the relative variances $\mathbb{S}[V_i]$ and $\mathbb{S}[W_i]$ of all paired-product estimators $V_i$ and $W_i$ are at most $e^2$. We rewrite Eq.~\eqref{eq:relvar_e2} as
\begin{align}
    \label{eq:promiseofsvvschedule}
    f\left( \frac{\beta_i + \beta_{i+1}}{2}\right) \ge \frac{f(\beta_i) + f(\beta_{i+1})}{2} - 1,
\end{align}
where $f(\beta)=\ln\big(\Zp(\beta)\big)$. It is important that $f(\beta)$ is a strictly decreasing convex function, which is the case if the Hamiltonian is non-negative. 

\begin{theorem}[Perfectly-balanced schedule length [Analogous to \cite{vstefankovivc2009adaptive}]]
\label{thm:promiseSVVschedule}
There exists a sequence $\beta_0<\cdots<\beta_\ell$ with $\beta_0=\beta_{\min}$ and $\beta_\ell=\beta_{\max}$ satisfying the condition in~\eqref{eq:promiseofsvvschedule} with equality and having length $\ell$ bounded from above by:
\begin{equation}
\label{eq:upperboundonlength}
    \ell \le \sqrt{\big(f(\beta_{\min})-f(\beta_{\max})\big) \cdot
    \frac{1}{2} \ln\left( \frac{f'(\beta_{\min})}{f'(\beta_{\max})} \right)}.
\end{equation}
\end{theorem}

The proof of Theorem~\ref{thm:promiseSVVschedule} can be found in Appendix~\ref{appendix:schedule_length}.

\begin{corollary}[Upper bound on perfectly-balanced schedule length] \label{cor:promiseSVVlength} The length $\ell$ of a perfectly-balanced schedule, as in Theorem~\ref{thm:promiseSVVschedule}, satisfies $    \ell \le \sqrt{q\cdot \ln n}$. 

\end{corollary}
\begin{proof}
By Theorem~\ref{thm:promiseSVVschedule}, the length of the schedule satisfies Eq.~\eqref{eq:upperboundonlength}.  We can further upper bound~$\ell$ in Eq.~\eqref{eq:upperboundonlength} as follows: observe that $f(\cdot)=\ln(\Zp(\cdot))$, so we have 
$$
f(\beta_{\min})-f(\beta_{\max})=\ln (\Zp(\beta_{\min})/\Zp(\beta_{\max})) \le \ln|\Omega| = q.
$$
The above inequality holds because $|\Omega| \geq Z(\beta) \geq 1$ for all $\beta \geq 0$.
Moreover, 
\begin{align*}
   f'(\beta) =  \Big( \ln \big( \Zp(\beta) \big) \Big)' = 
   \frac{\Zp'(\beta)}{\Zp(\beta)} =-\sum_{x\in\Omega} \mu_\beta(x) H(x).
\end{align*}
We have: 
$$
\frac{f'(\beta_{\min})}{f'(\beta_{\max})}= \frac{-f'(\beta_{\min})}{-f'(\beta_{\max})}\leq \frac{e-1}{e} \cdot n,
$$
where we used the fact that $-f'(\beta_{\min})\leq n$ (which follows from the equation above and the assumption that $H(x)\leq n$) and $-f'(\beta_{\max}) \ge -f'(\ln |\Omega|)\ge\frac{e}{e-1}$ (see Ref.~\cite[Eq.~35]{vstefankovivc2009adaptive}). 
\end{proof}

Theorem \ref{thm:promiseSVVschedule} establishes existence of a perfectly-balanced cooling schedule.

Determining temperatures satisfying Eq.~\eqref{eq:promiseofsvvschedule} as equality are challenging -- we instead work with {\em well-balanced} cooling schedules, which~satisfy:
\begin{equation}
  \label{eq:slowlyvaryingcond}
    c_1 \le \frac{\Zp(\beta_i) \Zp(\beta_{i+1})}{\Zp(\bar{\beta}_{i,i+1})^2} \le c_2, \; \forall i\in\{0,\ldots,\ell-1\},
\end{equation}
for suitably chosen constants $c_1, c_2$. The upper bound $c_2$ ensures that the relative variances $\mathbb{S}[V_i]$ and $\mathbb{S}[W_i]$ are not too large, while the lower bound $c_1$ ensures that the temperatures increase rapidly enough, so that the schedule is short. We will also call a cooling schedule that satisfies:
\begin{align}
\frac{\Zp(\beta_i) \Zp(\beta_{i+1})}{\Zp(\bar{\beta}_{i,i+1})^2} \le c_2, \; \forall i\in\{0,\ldots,\ell-1\},\end{align} a $c_2$-\emph{slowly varying} cooling schedule. 

We describe an iterative algorithm that,
given a cooling schedule ending at $\beta_i$, finds $\beta_{i+1}$ so that the condition in Eq.~\eqref{eq:slowlyvaryingcond} is satisfied. To achieve this, we estimate the ratios $\frac{\Zp(\beta_i)}{\Zp(\bar{\beta}_{i,i+1})}$ and $\frac{\Zp(\beta_{i+1})}{\Zp(\bar{\beta}_{i,i+1})}$ and use their product as the estimate for the relative variances $\mathbb{S}[V_i]$ and $\mathbb{S}[W_i]$. Obtaining highly accurate estimates of these ratios is difficult: indeed, if one could compute $\frac{\Zp(\beta_i)}{\Zp(\beta_{i+1})}$ with high accuracy, then one could follow a cooling schedule, estimate the products $\frac{\Zp(\beta_i)}{\Zp(\beta_{i+1})}$ for all $i\in \{0,\ldots,\ell\}$, multiply them and solve the original problem of estimating $\Zp(\beta_\ell)$.  
Fortunately, to decide if the well-balanced condition holds, it suffices to have ``crude'' estimates obtained by taking less samples.  To prove this, we use several results from Ref.~\cite{vstefankovivc2009adaptive}.
The following lemma bounds the number of large temperature increases in any cooling schedule by comparing it to a perfectly-balanced cooling schedule.

\begin{corollary}[Analogous to {\cite[Corollary~4.4]{vstefankovivc2009adaptive}}]
  \label{lem:varyingschedulebound}
Let  $\beta_{\min}=\gamma_0 < \cdots < \gamma_m=\beta_{\max}$ be an arbitrary cooling schedule such that:
   \begin{equation}
     \label{eq:varlb}
     \frac{\Zp(\gamma_j) \Zp(\gamma_{j+1})}{\Zp\big(\frac{\gamma_j + \gamma_{j+1}}{2}\big)^2} \ge e^2,
  \end{equation}
  for every $j=0,\dots,m-1$. 
Then, $m\le \ell$, where $\ell$ is the length of a perfectly-balanced cooling schedule $\beta_{\min}=\beta_0<\cdots<\beta_\ell=\beta_{\max}$ as in Theorem~\ref{thm:promiseSVVschedule}.
\end{corollary}
\begin{proof}
 Theorem~\ref{thm:promiseSVVschedule}  upper-bounds the length of a perfectly-balanced cooling schedule, which has relative variances equal to $e^2$. We show that a cooling schedule $\gamma_0<\cdots <\gamma_m$ with relative variances at least $e^2$ cannot be longer than the upper bound derived in Theorem~\ref{thm:promiseSVVschedule}. This requires showing that increasing the relative variance can only make the schedule shorter. Define:
  \begin{equation*}
    g(x, y) = \ln \frac{\Zp(x) \Zp(y)}{\Zp\left(\frac{x+y}{2}\right)^2} = 
    f(x) + f(y) - 2 f\left(\frac{x+y}{2}\right).
  \end{equation*}
  $g(x, y)$ is decreasing in $x$ for $x < y$~because
  \begin{equation*}
    \frac{\partial g(x,y)}{\partial x} = f'(x) - f'\left(\frac{x+y}{2}\right) < 0
  \end{equation*}
  due to strict convexity of $f$ and $x< (x+y)/2$.  Similarly one can show that $g(x, y)$ is increasing in~$y$ for $x < y$.
  We now show by induction that $\beta_j \le \gamma_j$ for every $j$. This is trivial for $j =
  0$. 
  By assumption
  $g(\beta_j, \beta_{j+1}) = 2$ because the $\beta$ schedule is
  perfectly-balanced, and $g(\gamma_j, \gamma_{j+1}) \ge 2$ from
Eq.~\eqref{eq:varlb}. Furthermore, $\beta_j \le \gamma_j$ by the
  induction hypothesis. Assume to the contrary
  that $\beta_{j+1} > \gamma_{j+1}$.  In this case, the monotonicity 
 of $g$ (in $x$,$y$ individually) implies that $g(\beta_j,\beta_{j+1}) > 
  g(\gamma_j, \gamma_{j + 1})$. But this is a contradiction since $g(\beta_j,\beta_{j+1})=2$ and  $g(\gamma_j, \gamma_{j + 1})> 2$. 
  This concludes the induction step $\beta_{j+1} \le \gamma_{j+1}$.
\end{proof}
Theorem~\ref{thm:promiseSVVschedule}, Corollary~\ref{cor:promiseSVVlength} and Corollary~\ref{lem:varyingschedulebound} upper-bounds the length of a set of well-balanced cooling schedules with $c_1 = e^2$ by $\sqrt{q \ln n}$.

\begin{definition}
\label{def:heavy}
Let $I = [b, c] \subseteq \{0,\dots,n\}$ and $h \in (0,1)$. We say that $I$ is
  {\em $h$-heavy} for $\beta>0$ if 
  $$
  \Pr_{X \sim
  \mu_\beta}[ H(X) \in I ] \ge h.
  $$
  The quantity $\Pr_{X \sim
  \mu_\beta}[ H(X) \in I ]$ is also called the weight of $I$ at inverse temperature $\beta$.
\end{definition}

\begin{lemma}[{\cite[Lemma 5.3]{vstefankovivc2009adaptive}}]
  \label{lem:hheavyinterval}
  Let $I = [b, c] \subseteq \{0,\dots,n\}$, $h \in (0,1)$. If  $I$ is $h$-heavy for two temperatures $\beta_1 < \beta_2$, then $I$ is also $h$-heavy for all temperatures $(1-\tau)\beta_1 + \tau \beta_2$, where~$\tau\in[0,1]$, i.e., the set of temperatures for which $I$ is $h$-heavy is a (possibly empty) subinterval of $[\beta_{\min}, \beta_{\max}]$.
\end{lemma}

Lemma \ref{lem:ratioest} shows that if we find some interval $I=[b,c]$ that is both sufficiently narrow 
and is $h$-heavy for two nearby
temperatures $\beta_1$ and~$\beta_2$, we can use that interval to estimate the ratio $\Zp(\beta_2)/\Zp(\beta_1)$:

\begin{lemma}[{\cite[Lemma~5.8]{vstefankovivc2009adaptive}}]
  \label{lem:ratioest}
Let $I = [b, c] \subseteq  \{0,\dots,n\}$, $\delta \in (0, 1]$ and $h\in (0,1)$. Suppose that
    $I$ is $h$-heavy for $\beta_1, \beta_2>0$ satisfying: 
    \begin{equation}\label{eq:maxinscrease}
      |\beta_1 - \beta_2| (c-b) \le 1.
    \end{equation}
    For $k\in \{1,2\}$, let $X_k \sim \mu_{\beta_k}$ and let $Y_k$ be the indicator function for the event $[H(X_k) \in I]$. Let $s = \ceil{(8/h) \cdot \ln (1/\delta)}$ and $U_k$ be the average of $s$ independent samples from $Y_k$. Let
    \begin{equation}
      \label{eq:estdef}
      \textsc{Est}(I, \beta_2, \beta_1) := \frac{U_1}{U_2} \exp(b(\beta_1 - \beta_2)).
    \end{equation}
    Then, with probability at least $1-4\delta$, we have:
    \begin{equation*}
      \label{eq:estproperty}
      \frac{1}{4e} \frac{\Zp(\beta_2)}{\Zp(\beta_1)} \le \textsc{Est}(I, \beta_2, \beta_1) \le 4e \frac{\Zp(\beta_2)}{\Zp(\beta_1)}.
    \end{equation*}
\end{lemma}
Lemma \ref{lem:ratioest} provides a way to compute $\Zp(\beta_i)/\Zp(\bar{\beta}_{i,i+1})$ and $\Zp(\beta_{i+1})/\Zp(\bar{\beta}_{i,i+1})$: find an interval which is $h$-heavy for both $\beta_i$ and $\beta_{i+1}$, and use Eq.~\eqref{eq:estdef} to estimate the two ratios up to a relative error of $4e$. This is possible thanks to  Lemma~\ref{lem:hheavyinterval}, ensuring that the midpoint $\bar{\beta}_{i,i+1}$ between~$\beta_i$ and~$\beta_{i+1}$ satisfies the assumptions of Lemma~\ref{lem:ratioest}. To generate a short cooling schedule, given some~$\beta_i$ we need to determine some $\beta_{i+1}$ such that: (i) we have an interval $I$ which is $h$-heavy for both $\beta_i$ and $\beta_{i+1}$; (ii) the well-balanced condition Eq.~\eqref{eq:slowlyvaryingcond} holds. The following lemma is used to determine if a given interval is $h$-heavy at a given temperature. 

\begin{lemma}[{\cite[Lemma~5.5]{vstefankovivc2009adaptive}}]
  \label{lem:isheavy}
 Let $I = [b, c] \subseteq \{0,\dots,n\}$, $\beta >0$ and  $\delta, h \in (0, 1]$. Let $X \sim \mu_\beta$ and let $Y$ be the indicator function for the event $[H(X)\in I]$. Let $s = \ceil{(8/h) \ln (1/\delta)}$ and~$U$ be the average of $s$ independent samples from $Y$. Define
    \begin{equation*}
      \textsc{IsHeavy}(I, \beta) = 
      \begin{cases} 
            \textsf{true}  & \textrm{if } U \ge 2h,\\ 
            \textsf{false} & \textrm{if } U <   2h. 
        \end{cases}
    \end{equation*}
     If $I$ is not $h$-heavy at $\beta$ we have $\Pr[\textsc{IsHeavy}(I, \beta) = \textsf{true}] \le \delta$ (where the probability is over the randomness in sampling from $\mu_\beta$), and if $I$ is $4h$-heavy at $\beta$ we have $\Pr[\textsc{IsHeavy}(I,\beta) = \textsf{false}] \le \delta$.
\end{lemma}
\begin{proof}
Assume that $I$ is $4h$-heavy. The expected number of samples that fall into $I$ is $\geq 4hs$. By Chernoff bound $I$ will receive  $\leq 2hs$ samples with probability at most $\delta$, i.e., $\Pr[U \le 2h] \le e^{-sh/8} \le \delta$. Now assume that $I$ is not $h$-heavy. The expected number of samples that fall into $I$ is $\leq hs$. Using Chernoff bound, $I$ will receive more than $2hs$ samples with probability at most $\delta$, i.e., $\Pr[U \ge 2h] \le e^{-sh/8} \le \delta$.
\end{proof}

\begin{corollary}
  \label{cor:findheavy}
Let $\beta>0$, $P$ be an arbitrary partition of $\{0,\dots,n\}$, and $h = \frac{1}{k|P|}$ for $k \ge 4$. Let~$F$ be a (possibly empty) subset of $P$ such that no interval in $F$ is $\frac{1}{|P|}$-heavy at $\beta$. Suppose we obtain $s =  \ceil{(8/h) \ln (1/\delta)}$ samples from $\mu_{\beta}$ and select an interval in $P \setminus F$ that received the largest number of samples. With probability at least $1 - \delta |P|$, this interval is $h$-heavy at~$\beta$.
\end{corollary}

\begin{proof}
Observe that there exists a $kh=(1/|P|)$-heavy interval in $P \setminus F$: if this were not the case, then every interval in the partition $P$ would have weight $<1 / |P|$ (since intervals in $F$ also have weight $< 1 / |P|$) and the overall weight would be $<1$ which is a contradiction. 
    Furthermore, $k \ge 4$, therefore there is at least one $4h$-heavy interval in $P \setminus F$.  
  We now apply the union bound for the following event: every $4h$-heavy interval receives at least $2hs$ samples, and every interval that is not $h$-heavy receives less than $2hs$ samples. This event occurs with probability at least $1-|P|\cdot \delta$, and conditioned on this event, the interval returned by the algorithm (i.e., the one that received the largest number of samples) is $h$-heavy.
\end{proof}

We use the following special partition $P$ in our schedule generation algorithm, that provides suitable candidate intervals $I$ from which one can select an $h$-heavy interval: 
\begin{itemize}
\item Set $P \leftarrow \emptyset$, $b \leftarrow 0$.
\item Repeat until $b \ge n$: add the interval $\{b, \dots, b +
  \floor{b/\sqrt{q}}\}$ to $P$; set $b \leftarrow b+\floor{b/\sqrt{q}}+1$. 
\item Return the set of intervals $P$.
\end{itemize}
\begin{lemma}[{\cite[Lemma~5.1]{vstefankovivc2009adaptive}}]
  \label{lem:partitionlength}
  $P$ has
  size $|P| \le 4 \sqrt{q} \ln n$.
\end{lemma}

 The last ingredient is the binary search described in Algorithm \ref{alg:binarysearch}.

\begin{algorithm} 
  \SetAlgoLined
  \SetKwInOut{Input}{input}\SetKwInOut{Output}{output}
  
  \Input{ Monotone predicate ${\cal P}$,\footnotemark{} interval $[a,b]$ such that ${\cal P}(a) =\textsf{true}$, precision $\alpha$.}
  \Output{ $b$ if ${\cal P}(b) =$ \textsf{true}, otherwise an $x$ such that ${\cal P}(x) =$ \textsf{true} and ${\cal P}(x + \alpha) =$ \textsf{false}}
  \If{${\cal P}(b)$}{\Return $b$}
  Set $\lambda \leftarrow a, \rho \leftarrow b$\;
  \While{$\rho - \lambda > \alpha$}{
    \eIf{${\cal P}(\frac{\lambda + \rho}{2})$}{ 
    $\lambda \leftarrow \frac{\lambda + \rho}{2}$
    }{
    $\rho \leftarrow \frac{\lambda + \rho}{2}$
    }
  }
  \Return $\lambda$
  \caption{Binary search subroutine, denoted $\textsc{BinarySearch}({\cal P}, [a,b], \alpha)$.}
  \label{alg:binarysearch}
\end{algorithm}
\footnotetext{A \emph{monotone predicate} $\cal P$ is a Boolean function defined on a totally ordered set with the following property: if ${\cal P}(x) = \textsf{true}$, then ${\cal P}(y) = \textsf{true}$ for all $y \le x$ in the domain, i.e., $\cal P$ is monotone decreasing.}

\begin{theorem}
\label{thm:findingSVVschedule}
There exists a procedure that, with probability at least $1-\delta$,  computes a $2 \cdot 10^5$-slowly-varying schedule with length at most $11\sqrt{q} \ln n$, and uses at most 
$
5\cdot 10^4 \cdot q \ln^2 n \cdot (\ln q + \ln n)^2 \ln(1/\delta) 
$
samples from the Gibbs distribution.
\end{theorem}
\begin{proof}
See Algorithm~\ref{alg:cschedgen}.
\begin{algorithm}
  \SetAlgoLined
  \SetKwInOut{Input}{input}\SetKwInOut{Output}{output}
  
  \Input{Initial temperature $\beta_0$, partition $P$, largest temperature $\beta_{\max}$, probability $\delta$.}
  \Output{Set of temperatures $\beta_0,\dots,\beta_k=\beta_{\max}$.}
  Set $h \leftarrow \frac{1}{8 |P|}$ throughout the algorithm and subroutine calls\;
  Set $k \leftarrow 0$, $F \leftarrow \emptyset$\;
  \While{$\beta_k < \beta_{\max}$}{
    Apply the procedure of Corollary~\ref{cor:findheavy} with $\beta = \beta_k$, $h = \frac{1}{8 |P|}$, and $F$ as the set of forbidden intervals; let its return value be the interval $I = \{b,\dots,c\}$\; 
    Set $L \leftarrow \min \{\beta_k + 1/(c-b), q\}$, with the convention $1/0 = \infty$\;
    Compute $L^* \leftarrow \textsc{BinarySearch}(\textsc{IsHeavy}(I, \cdot) = \textsf{true}, [\beta_k, L], 1/2n)$\;
    Compute $\beta^* \leftarrow \textsc{BinarySearch}(\textsc{Est}(I, \frac{\cdot + \beta_k}{2}, \beta_k)\textsc{Est}(I, \cdot, \frac{\cdot + \beta_k}{2}) \le 1500, [\beta_k, L^*], 1/2n)$\;
    If $\beta^* = L^* < L$, set $F \leftarrow F \cup \{I\}$\;
    Set $\beta_{k+1} \leftarrow \beta^*, k \leftarrow k+1$\;
  }
  \Return $\beta_1,\dots,\beta_k,\beta_{\max}$
\caption{Classical schedule generation procedure. The partition $P$ is generated according to Lemma \ref{lem:partitionlength}.}
\label{alg:cschedgen}
\end{algorithm}
\footnotetext{We use $(\cdot)$ as the argument for inline function definitions; e.g., the expression $\textsc{BinarySearch}(\textsc{IsHeavy}(I, \cdot) = \textsf{true}, [\beta_k, L], 1/2n)$ means that the value of the predicate for binary search, ${\cal P}$, at a point $x$ is the expression $\textsc{IsHeavy}(I, x) = \textsf{true}$, i.e. ${\cal P}(x)=1 \Leftrightarrow \textsc{IsHeavy}(I, x) = \textsf{true}$.}

We prove correctness of this algorithm and analyze its complexity. Fix 
   $h := 1/(8 |P|)$. By Lemma~\ref{lem:partitionlength}, we have $|P| \le 4\sqrt{q} \ln n$. Using Corollary~\ref{cor:findheavy} in the first line of the ``while'' loop, we can always find an interval $I$ that is $h$-heavy for $\beta_k$ and does not belong to the set of forbidden intervals~$F$, provided that none of the forbidden intervals in $F$ is $8h=(1/|P|)$-heavy at $\beta_k$. This will be shown below. The interval $I$ depends on the current inverse temperature $\beta_k$. For now, we neglect the failure probability of the algorithm for determining $I$ in Corollary~\ref{cor:findheavy} as well as the failure probabilities of the two binary searches in the third and fourth lines of the ``while'' loop and account for these at the end of the proof by a union bound argument.  We will also show that the two predicates used inside the binary searches in the third and fourth lines are monotone and are satisfied at the left ends of their respective search intervals, as required for Algorithm~\ref{alg:binarysearch}.
  
  To bound the total number of temperatures in the final schedule, we analyze three mutually exclusive cases that can arise in each iteration: (1) $\beta^*=L^*=L$, (2) $\beta^*=L^*<L$, (3) $\beta^* < L^*$. Observe that in cases (2) and (3) the binary searches return inverse temperatures that are not equal to the right endpoint of their respective search intervals.  Therefore, if these inverse temperatures are increased by the precision $\alpha=1/(2n)$, the predicates are no longer satisfied. This allows us to bound the number of steps in cases (2) and (3).

\paragraph*{Case 1}  In case (1), we move by setting $L^* = L$ and $\beta^* = L^*$. These moves are called ``long moves'' in \cite{vstefankovivc2009adaptive} because the inverse temperature is increased by the maximally possible value according to the requirement \eqref{eq:maxinscrease} in Lemma~\ref{lem:ratioest}. To  bound the  number of long moves, we use the following lemma. 
  \begin{lemma}
\label{lem:longmoves}
The number of ``long moves'' in Algorithm~\ref{alg:cschedgen}, where we set $\beta^* = L^* = L$, is at most~$6 \sqrt{q} \ln n$.
\end{lemma}

We defer the proof of this lemma to Appendix~\ref{app:longmoveslemma} as it is rather long and technical.

\paragraph*{Case 2}  We move by setting $L^* < L$ and $\beta^* = L^*$ and show that when such a move takes place, the interval $I$ is not $8h=(1/|P|)$-heavy for any $\beta \ge L^*$. This ensures that we can correctly apply Corollary~\ref{cor:findheavy} in all subsequent iterations, which are executed with a set $F$ of forbidden intervals that now includes $I$ and only consider inverse temperatures $\beta\ge L^*$. The value~$L^*$ is obtained by applying $\textsc{BinarySearch}(\textsc{IsHeavy}(I, \cdot) = \textsf{true}, [\beta_k, L], 1/2n)$. Clearly, it holds that $\textsc{IsHeavy}(I, \beta_k)=\textsf{true}$. Thus, the predicate used in the first binary search of Algorithm~\ref{alg:cschedgen} is satisfied at the left endpoint $\beta_k$ of the search interval. This predicate is monotone, which is implied by Lemma~\ref{lem:hheavyinterval}.
    
    The fact $L^*<L$ means that the value returned by binary search is not equal to the right end of the search interval. Therefore, by the
    properties of the binary search procedure we have $\textsc{IsHeavy}(I, L^*) = \textsf{true}$ and $\textsc{IsHeavy}(I, \rho) = \textsf{false}$ for some $\rho$ with $L^* < \rho \le L^* + 1/(2n)$. By Lemma~\ref{lem:isheavy}, $I$ is $4h$-heavy for $L^*$ but is not $4h$-heavy anymore for $\rho$. 
    
    We first prove that $I$ is not $8h$-heavy at any inverse temperature in the interval $[L^*, \rho]$.
    To this end, we bound the weight of $I$ at $\beta'$, for any $\beta' \in [L^*, \rho]$, as follows:
\begin{equation*}
\begin{aligned}
    \frac{1}{\Zp(\beta')} \sum_{x : H(x) \in I} e^{-\beta' H(x)} 
    &\le \frac{1}{\Zp(\rho)} \sum_{x : H(x) \in I} e^{-\rho H(x)} e^{1/2} \le 4h e^{1/2} < 8h.
    \end{aligned}
\end{equation*}
In the chain of inequalities above, we used the following facts: for the first inequality, $\Zp(\beta') \ge \Zp(\rho)$ because $\Zp(\cdot)$ is non-increasing, and $\beta' \ge \rho - 1/(2n)$; for the second inequality we used $H(x) \le n$; for the third inequality, we used the fact that the weight of $I$ at $\rho$ is at most $4h$ (recall Definition~\ref{def:heavy}).   

It remains to prove that $I$ is not $8h$-heavy for any $\beta'>\rho$.  Assume to the contrary that such a $\beta'$ exists. This would automatically imply that $I$ is $4h$-heavy at $\beta'$. Combined with Lemma~\ref{lem:hheavyinterval}, this would imply that $I$ is $4h$-heavy for all temperatures inside the interval $[L^*,\beta']$. But this contradicts that $I$ is not $4h$-heavy for $\rho$.

We can therefore apply Corollary~\ref{cor:findheavy} in subsequent iterations where~$I$ is forbidden. Because~$I$ is added to the set $F$ of forbidden intervals, an iteration in which we set $\beta^* = L^*$ with $L^* < L$ can only take place at most once per interval $I$. Hence, the number of these iterations is at most $|P| \le 4 \sqrt{q} \ln n$.

\paragraph*{Case 3} In case (3), we move by setting $\beta^* < L^*$.
  Notice that 
  $$\textsc{Est}(I, \frac{\beta_k + \beta_k}{2},
  \beta_k)\allowbreak \textsc{Est}(I, \beta_k, \frac{\beta_k +
    \beta_k}{2}) \le 16e^2 \le 1500
    $$ by Lemma \ref{lem:ratioest}. Thus, the
  predicate used in the second binary search  of Algorithm~\ref{alg:cschedgen} is satisfied at the left endpoint $\beta_k$ of the search interval. Moreover, this predicate is monotone, which follows from the discussion in the proof of Corollary~\ref{lem:varyingschedulebound}.
  Hence, binary search with precision $1/2n$ determines a
  value $\lambda$ such that there exists $\rho$ satisfying:
  \begin{align*}
  \small
    \textsc{Est}\Big(I, \frac{\lambda + \beta_k}{2},
    \beta_k\Big)\textsc{Est}\Big(I, \lambda, \frac{\lambda + \beta_k}{2}\Big) &\le
    1500, \\  
    \textsc{Est}\Big(I, \frac{\rho + \beta_k}{2},
    \beta_k\Big)\textsc{Est}\Big(I, \rho, \frac{\rho + \beta_k}{2}\Big) &>
    1500, \\
    \rho - \lambda &\le 1/2n, \\   \lambda &\ge \beta_k.
  \end{align*}
  Using Lemma \ref{lem:ratioest}, we have:
  \begin{equation}
   \small
  \begin{aligned}
  \label{eq:ratio1}
    \frac{\Zp(\beta_k) \Zp(\lambda)}{\Zp(\frac{\beta_k + \lambda}{2})^2} &\le 16e^2 \textsc{Est}\Big(I, \frac{\lambda + \beta_k}{2},
    \beta_k\Big)\textsc{Est}\Big(I, \lambda, \frac{\lambda + \beta_k}{2}\Big) \le 2 \cdot 10^5.
  \end{aligned}
  \end{equation}
  Furthermore,
  \begin{equation}
      \small
  \begin{aligned}
  \label{eq:ratio2}
    \frac{\Zp(\beta_k) \Zp(\rho)}{\Zp(\frac{\beta_k + \rho}{2})^2} &\ge \frac{1}{16e^2} \textsc{Est}\Big(I, \frac{\rho + \beta_k}{2},
    \beta_k\Big)\textsc{Est}\Big(I, \rho, \frac{\rho + \beta_k}{2}\Big) \ge \frac{1500}{16e^2},
  \end{aligned}
 \end{equation}
  because the predicate is false at $\rho$. It remains to lower bound the  LHS of Eq.~\eqref{eq:ratio1}. In order to bound the ratio, first observe that $\Zp(\lambda)
  \ge \Zp(\rho)$. Then, observe that for every $\varepsilon > 0$, the following inequality holds:
  \begin{align*}
     \Zp(\beta) \geq \Zp(\beta +\varepsilon) &= \sum_x e^{-(\beta + \varepsilon) H(x)} \geq e^{-n \varepsilon}\cdot \Zp(\beta).
   \end{align*}
  It then follows that $\Zp(\frac{\beta_k + \lambda}{2}) \le
  \Zp(\frac{\beta_k + \rho}{2})e^{1/4}$, because $\frac{\beta_k + \rho}{2} - \frac{\beta_k + \lambda}{2} \le \frac{1}{4n}$.
  The LHS of Eq.~\eqref{eq:ratio1} can be lower bounded by
  \begin{equation}
    \label{eq:varlbproof}
    \frac{\Zp(\beta_k) \Zp(\lambda)}{\Zp(\frac{\beta_k + \lambda}{2})^2} \ge  \frac{\Zp(\beta_k) \Zp(\rho)}{e^{1/2} \Zp(\frac{\beta_k + \rho}{2})^2} \ge \frac{1500}{16e^{2.5}} \ge 10 \ge e^2,
  \end{equation}
  where the first inequality used $\Zp(\frac{\beta_k + \lambda}{2}) \le
  \Zp(\frac{\beta_k + \rho}{2})e^{1/4}$ and the second inequality used Eq.~\eqref{eq:ratio2}.  
  
  We now conclude the proof. We showed above that, in the iterations for which $\beta^* < L^*$, we select inverse temperatures that satisfy condition \eqref{eq:varlbproof}. The number of such inverse temperatures is at most $\sqrt{q \ln n}$ by Corollary~\ref{lem:varyingschedulebound}. We have also shown that the number of iterations in which we set $\beta^* = L^* = L$ is at most $6\sqrt{q} \ln n$ using Lemma~\ref{lem:longmoves}, and the number of iterations in which se set $\beta^* = L^* < L$ is at most $4\sqrt{q} \ln n$. Therefore, the length of the computed schedule is at most $6 \sqrt{q} \ln n + 5\sqrt{q \ln n} \leq 11\sqrt{q} \ln n$.
  
  In each step we perform binary search
  (twice) with precision $1/(2n)$ over a domain that is contained in
  $[0, \beta_{\max}]$; the total number of binary search iterations per step
  is at most $\log (4n \beta_{\max}) \le 8 (\ln \beta_{\max} + \ln
  n)$. Each binary search iteration requires at most $2s$ samples,
  where $s=(8/h)\cdot \ln (1/\delta') = 64 |P| \ln (1/\delta')$ is given by Lemma~\ref{lem:ratioest}. To ensure that the algorithm is successful with probability at least $1-\delta$, we apply a union bound over  $T=88\sqrt{q} \ln n (\ln \beta_{\max} + \ln n)$ subroutine calls, and we choose $\delta' = \delta\cdot T^{-1}$ for each subroutine. 

  We obtain a schedule with length
  $11\sqrt{q} \ln n$ by taking at most $2sT \le 5\cdot 10^4 q \ln^2 n (\ln q + \ln n) \ln (1/\delta')$ samples from the Gibbs distribution $\mu_{\beta}$.  
  Substituting the value for~$\delta'$ and simplifying by using the assumptions discussed at the beginning of Section \ref{Sec:Classical}, plus the assumption $\ln n \ge 5 + \ln (\ln q + \ln n) + \ln \ln n$, the total number of samples is at most $5\cdot 10^4 q \ln^2 n (\ln q + \ln n)^2 \ln (1/\delta)$.
\end{proof}

\subsection{Classical algorithm}
\begin{theorem}
\label{thm:classicalmainthm}
Let $n\geq 1$, $\varepsilon\in (0,1)$, and let $0\leq \beta_{\min}<\beta_{\max}$. Let $H:\Omega\rightarrow \{0,\dots,n\}$ be a Hamiltonian,  $Q=\Zp(\beta_{\max})/\Zp(\beta_{\min})$ and $q=\ln |\Omega|$. There exists a classical algorithm that satisfies the following: with probability at least $4/5$, the algorithm uses $\tilde{O}\big(q \cdot \varepsilon^{-2}\big)$ Gibbs distribution samples and outputs $\widehat{Q}$ that approximates $Q$ up to relative error $\varepsilon$ 

\end{theorem}

\begin{proof}
The proof has the following structure: 
\begin{enumerate}
    \item Produce a cooling schedule schedule $\{\beta_1,\ldots,\beta_\ell\}$.
    \item Bound relative variance of $\lbrace W_i,V_i \rbrace$.  
    
    \item Estimate $\widehat{W_i}\approx {W}_i,\widehat{V_i}\approx {V}_i$ and output $\prod_i \widehat{W}_i/\prod_i \widehat{V}_i$. 
\end{enumerate}

Step $(1)$ uses Theorem~\ref{thm:findingSVVschedule} to find a slowly-varying sequence of inverse temperatures. To see that the variables $V_i,W_i$ in step $(2)$ have bounded variance, observe that:
\begin{equation}
\small
\begin{aligned}
    \varS[W_i]&=\varS[V_i] = \frac{\Zp(\beta_i)\Zp(\beta_{i+1})}{\Zp\left(\frac{\beta_i + \beta_{i+1}}{2} \right)^2} \\ &= \exp\Big(-2f\left( \frac{\beta_i + \beta_{i+1}}{2}\right) + f(\beta_i) + f(\beta_{i+1})\Big) \leq 2\cdot 10^5,
\end{aligned}
\end{equation}
where the last inequality used that the schedule from  Theorem~\ref{thm:findingSVVschedule} is  $2\cdot 10^5$-slowly-varying.  Furthermore, 
$$
\frac{\prod_i \mathbb{E}[V_i]}{\prod_i \mathbb{E}[W_i]}= \frac{\Zp(\beta_{\max})}{\Zp(\beta_{\min})} =Q.
$$
From Theorem~\ref{thm:dyerfrieze},
obtaining $4\cdot 10^5\cdot \ell/(\eta \bar{\varepsilon}^2)$ samples from $W_i$ and $V_i$ suffices to produce $\widehat{W},\widetilde{V}$, such that:    \begin{align*}
\small
&\Pr\left[
(1-\bar{\varepsilon})\cdot \prod_{i}\E[W_i] \le \widehat{W} \le (1+\bar{\varepsilon})\cdot \prod_{i}\E[W_i]
\right]
\ge 1-\eta ,
\end{align*}
for a given $\bar{\varepsilon} > 0$ (and similarly for $\hat{V}$).
With probability at least $1-2\eta$, we have:
    \begin{align*}
    \small
    (1 - 3\bar{\varepsilon}) \cdot Q &\le
    (1-\bar{\varepsilon})^2\cdot Q  =  (1-\bar{\varepsilon})^2\cdot \frac{\prod_i \mathbb{E}[V_i]}{\prod_i \mathbb{E}[W_i]} \\ &\le 
    \widehat{V} / \widehat{W} \leq  (1+\bar{\varepsilon})^2\cdot \frac{\prod_i \mathbb{E}[V_i]}{\prod_i \mathbb{E}[W_i]}\leq (1 + 3\bar{\varepsilon}) \cdot Q 
    \end{align*}
Choosing $\bar{\varepsilon} = \varepsilon/3$ shows that the proposed scheme correctly outputs a $\varepsilon$-approximation of $Q$. From Theorem~\ref{thm:findingSVVschedule}, 
$5 \cdot 10^4\cdot  q \ln^2n (\ln q + \ln n)^2 \ln (1/\delta) $
  samples suffice to find a slowly-varying schedule $\{\beta_1.\ldots,\beta_\ell\}$, where~$\delta$ is the maximum failure probability of the schedule generation algorithm. Choose $\delta = 1/10$. For the application of Theorem.~\ref{thm:dyerfrieze}, choose $\eta = 1/20$;
  the complexity of obtaining the estimates $\widehat{W},\widehat{V}$ is
$$
2\ell\cdot (4\cdot 10^5\cdot \ell/(\eta \bar{\varepsilon}^2)) \leq 5 \cdot 10^7 \ell^2  / \varepsilon^2 \le 7\cdot 10^{9} q \ln^2 n / \varepsilon^2,
$$
where we used the expression for $\ell=11\sqrt{q}\ln n$ given in Theorem~\ref{thm:findingSVVschedule}. 
Hence, with probability at least $4/5$
, our classical algorithm  produces an $\varepsilon$-relative estimator of $Q$ using 
$$
5 \cdot 10^5\cdot  q \ln^2n \cdot (\ln q + \ln n)^2 + 7\cdot 10^{9} \cdot q \ln^2 n\cdot  \varepsilon^{-2}
$$
Gibbs distribution samples.
\end{proof}

\section{Quantum algorithm}
\label{sec:quantum} 
\subsection{Cooling schedule, quantumly}
Our quantum algorithm is much simpler than the classical algorithm.
For the schedule generation, the variances of the random variables $\mathbb{S}[V_i]$ and $\mathbb{S}[W_i]$ can be estimated with quantum amplitude estimation thanks to a simple relationship between the variances of these random variables and the overlap between the qsamples $|\mu_{\beta_i}\>$ and~$|\mu_{\beta_{i+1}}\>$.

\begin{fact}[Variance and overlap]
The overlap of Gibbs qsamples can be written as 
\label{fact:variances_overlap}
\begin{equation}
\begin{aligned}\label{eq:large_overlap_2}
    \left| \langle \mu_{\beta_i} | \mu_{\beta_{i+1}} \rangle \right|^2
    &=
    \left| \sum_{x\in\Omega} \sqrt{\frac{{e^{-\beta_i H(x)}}}{\Zp(\beta_i)}} \cdot \sqrt{\frac{{e^{-\beta_{i+1} H(x)}}}{\Zp(\beta_{i+1})}} \right|^2 \\ 
    &=
    \frac{\Zp(\bar{\beta}_{i,i+1})^2}{\Zp(\beta_i) \cdot \Zp(\beta_{i+1})} 
    =
    \frac{1}{\mathbb{S}[W_i]} = \frac{1}{\mathbb{S}[V_i]},
\end{aligned}
\end{equation}
i.e., it equals the reciprocals of the variances $\mathbb{S}[V_i]$,~$\mathbb{S}[W_i]$.\footnote{It is now easy to see that, for $\mathbb{S}[X_i]=1/|\<\mu_{\beta_i}|\mu_{\gamma_{i,i+1}}\>|^2$ and $\mathbb{S}[V_i]=\mathbb{S}[W_i]=1/|\<\mu_{\beta_i}|\mu_{\beta_{i+1}}\>|^2$, the quantity $|\<\mu_{\beta_i}|\mu_\beta\>|$ is monotonically decreasing in $\beta$ (where $\beta\ge \beta_i$), hence $\mathbb{S}[V_i],\mathbb{S}[W_i]\leq \mathbb{S}[X_i]$.}
\end{fact}
We again aim to find a cooling schedule that satisfies:
$
  c_1 \le (\Zp(\beta_i) \cdot \Zp(\beta_{i+1}))(\Zp(\bar{\beta}_{i,i+1})^2)
  \le c_2,
$
where $c_1,c_2$ are constants (see Eq.~\eqref{eq:slowlyvaryingcond}). Using Fact~\ref{fact:variances_overlap}, this can be written~as
$
    1/c_2 \le |\< \mu_{\beta_i} | \mu_{\beta_{i+1}} \>|^2 \le 1/c_1.
$
We now apply the result below with $|\psi\>=|\mu_{\beta_i}\>$,  $P=|\mu_{\beta_{i+1}}\>\<\mu_{\beta_{i+1}}|$, and constant $\varepsilon$ to estimate $|\<\mu_{\beta_i}|\mu_{\beta_{i+1}}\>|^2$. The following is a corollary of \cite[Thm.~6]{harrow2020adaptive}.

\begin{corollary}[Non-destructive amplitude estimation]\label{cor:ndampest_additive}
Let $|\psi\>$ be an arbitrary quantum state and $P$ an arbitrary projector. Let $R_\psi=2|\psi\>\<\psi|-\id$. Let $\varepsilon\in (0,1)$. Then, there is a quantum algorithm $\A$ that starts in the initial state $|\psi\>$ and with probability at least $1-\eta$ outputs an estimate $\hat{p}$ of $p=\<\psi|P|\psi\>$ with additive error $\varepsilon$. Additionally, $\A$ restores the state $|\psi\>$ with probability $1-\eta$ and  invokes the controlled reflection $R_\psi$ $O\big( (1/\varepsilon) \cdot \ln( 1/\eta) \big)$ many times.
\end{corollary}

Let $|\phi\>$ and $|\psi\>$ be two arbitrary quantum states.  Define the projectors $P_\phi=|\phi\>\<\phi|$ and $P_\phi^\perp=\id-P$ for the state $|\phi\>$ and  similarly define $P_\psi$.

\begin{fact}\label{lem:jumpbymeasure}
Let $\ket{\phi},\ket{\psi}$ be quantum states. Assume that the transition probability $a=|\langle \phi\vert \psi\rangle |^2$ is bounded from below by some constant. 
Then, starting with~$|\phi\>$, we can prepare $|\psi\>$ with probability $1-\eta$, by performing at most $O\big(\ln(1/\eta)\big)$ measurements $\{P_\phi,P_\phi^\perp\}$ and $\{P_\psi,P_\psi^\perp\}$.
\end{fact}

\begin{proof}
The probability that we fail to prepare $\ket{\psi}$ after performing $2k+1$ measurements is given by $(1-a) \big(a^2 + (1-a)^2\big)^k$.
This result is easily established by observing that everything can be analyzed in the two dimensional space spanned by $P_\psi|\phi\>$ and $P_\psi^\perp|\phi\>$.  Consider the two bases $\{|\phi\>,|\phi^\perp\>\}$ and  $\{|\psi\>,|\psi^\perp\>\}$ for this subspace, where $\<\phi|\phi^\perp\>=0$ and $\<\psi|\psi^\perp\>=0$. The factor $(1-a)$ corresponds to the transition $|\phi\> \rightarrow |\psi^\perp\>$, i.e., that the first measurement fails to prepare $|\psi\>$. The term $a^2$ corresponds to the transition $|\psi^\perp\> \rightarrow |\phi^\perp\> \rightarrow |\psi^\perp\>$ and the term $1-a$ corresponds to the transition $|\psi^\perp\> \rightarrow |\phi\> \rightarrow |\psi^\perp\>$, i.e., that $P_\psi$ measurement followed by the $P_\phi$ measurement fails to prepare $|\psi\>$. It is clear that we can make the failure probability smaller than $\eta$ by choosing $k=O\big(\ln(1/\eta)\big)$ provided that $a$ is bounded from below by a constant.
\end{proof}

We now describe a quantum algorithm to obtain a schedule that matches Theorem~\ref{thm:promiseSVVschedule}. 
The difference between the classical and quantum setting is mainly due to the classical estimator in Eq.~\eqref{eq:relvar}, inaccurate compared to the quantum estimator. The quantum algorithm is more efficient than the classical algorithm, and improves the length quadratically in $\ln n$. 

\begin{theorem}[Quantum schedule generation]
\label{thm:qschedulegen}
With probability $\geq 1-\delta$,
Algorithm \ref{alg:qschedgen} computes
a $15$-slowly-varying schedule with length at most $\sqrt{q \ln n}$, and uses $O\Big(\sqrt{q \ln n}\big(\ln q + \ln n\big)\cdot \big(\ln q + \ln \ln n + \ln (1/\delta)\big) \Big)$ reflections about Gibbs states at different inverse~temperatures.
\end{theorem}

\begin{proof}

\begin{algorithm}
  \SetAlgoLined
  \SetKwInOut{Input}{input}\SetKwInOut{Output}{output}
  
  \Input{Initial temperature $\beta_0$, largest temperature $\beta_{\max}$, probability $\delta$, constant $c_2$.}
  \Output{Set of inverse temperatures $\beta_0,\dots,\beta_k=\beta_{\max}$.}
  Set $k \leftarrow 0$\;
  \While{$\beta_k < \beta_{\max}$}{
   Define the function $f_o(\beta) := |\<\mu_{\beta_k}|\mu_{\beta}\>|^2$, where the overlap is evaluated using Corollary~\ref{cor:ndampest_additive} 
   with additive error $0.005$\;
   Compute $\beta^* \leftarrow \textsc{BinarySearch}( f_o(\cdot) \ge 0.075, [\beta_k, q], 1/2n)$\;
    Set $\beta_{k+1} \leftarrow \beta^*, k \leftarrow k+1$\;
  }
  \Return $\beta_1,\dots,\beta_k,\beta_{\max}$
  \caption{Quantum schedule generation procedure.}
  \label{alg:qschedgen}
\end{algorithm}

  We first bound the number of iterations of the algorithm. Assume for now that all subroutines are successful, and we use a union bound at the end.
  Notice that $|\<\mu_{\beta_k}|\mu_{\beta_k}\>|^2 = 1$ so the
  predicate used in binary search is satisfied at the left endpoint of
  the interval $[\beta_k, q]$. If the binary search (as given in
  Alg.~\ref{alg:binarysearch}) returns $q$, we are
  done. Otherwise, it determines $\lambda, \rho$ such that:
  \begin{align*}
    |\<\mu_{\beta_k}|\mu_{\lambda}\>|^2 &\ge 0.075 - 0.005,\\
    |\<\mu_{\beta_k}|\mu_{\rho}\>|^2 &< 0.075 + 0.005, \\ 
    \rho - \lambda &\le 1/2n, \;     \lambda \ge \beta_k.
  \end{align*}
  We have
  \begin{align*}
    \frac{\Zp(\beta_k) \Zp(\lambda)}{\Zp(\frac{\beta_k + \lambda}{2})^2} = \frac{1}{|\<\mu_{\beta_k}|\mu_{\lambda}\>|^2} \le \frac{1}{0.075 - 0.005} \le 15,
  \end{align*}
  which shows that the schedule is $15$-slowly-varying. Also,
  \begin{align*}
    \frac{\Zp(\beta_k) \Zp(\rho)}{\Zp(\frac{\beta_k + \rho}{2})^2} = \frac{1}{|\<\mu_{\beta_k}|\mu_{\rho}\>|^2} \ge \frac{1}{0.075 + 0.005} = 12.5.
  \end{align*}
  It can be shown that $\Zp(\lambda)
  \ge \Zp(\rho)$, and that $\Zp(\frac{\beta_k + \lambda}{2}) \le
  \Zp(\frac{\beta_k + \rho}{2})e^{1/4}$. 
  
  Using these facts, we can write:
  \begin{equation}\label{eq:quantumvarlbproof}
    \frac{\Zp(\beta_k) \Zp(\lambda)}{\Zp(\frac{\beta_k + \lambda}{2})^2} \ge 
    \frac{\Zp(\beta_k) \Zp(\rho)}{e^{1/2} \Zp(\frac{\beta_k + \rho}{2})^2} \ge 
    e^{-1/2} \, 12.5\ge e^{2}.
  \end{equation}
  
{We see that the cooling schedule obtained by setting $\beta_{k+1}=\lambda$ in each iteration satisfies the condition in Corollary~\ref{lem:varyingschedulebound}, which coincides with the bound in (\ref{eq:quantumvarlbproof}).  This implies that the length of the resulting cooling schedule is bounded from above by the length of a perfectly balanced cooling schedule.  The latter in turn is bounded from above by $\sqrt{q \ln n}$, which is established in Corollary~\ref{cor:promiseSVVlength}.
}
 In each step we perform binary search with precision $1/2n$ over a domain that is contained in $[0,
    \beta_{\max}]$, which implies that the total number of binary search iterations per step is
  at most $\log (2n \beta_{\max}) \le 2 (\ln \beta_{\max} + \ln n)$. The total number of binary searches in all steps is therefore at most $2\sqrt{q \ln n}(\ln \beta_{\max} + \ln n)$.

  Each binary search invokes amplitude estimation subroutine in Corollary~\ref{cor:ndampest_additive} with additive error $\varepsilon$ set to the constant $0.005$. To ensure that the entire algorithm succeeds with probability at least $1-\delta$, we choose the maximum probability of failure
  $\eta = \delta / \big(4\sqrt{q \ln n}(\ln \beta_{\max} + \ln n)\big)$ for the
  amplitude estimation subroutine. By the union bound, the probability that at least one amplitude estimation subroutine fails is $\le \delta/2$.
  Each amplitude estimation requires 
  $O(\ln(1/\eta))$ many reflections since $\varepsilon$ is a constant.
  Hence, the total number of reflections for all binary searches is
  $O\big(\sqrt{q \ln n}(\ln \beta_{\max} + \ln n) \ln (1/\eta)\big)$, which, in terms of $\delta$ and using the assumption $\beta_{max} \le q$ (see the proof of Theorem~\ref{thm:findingSVVschedule}),~is
  \begin{equation*}
      O\big( \sqrt{q \ln n} (\ln q + \ln n)(\ln q + \ln \ln n  + \ln(1/\delta)) \big).
  \end{equation*}
  
  It remains to bound the time it takes to iteratively prepare the states 
  $\ket{\mu_{\beta_0}},\dots,\ket{\mu_{\beta_k}}$ in the schedule. 
  The overlap between all adjacent qsamples is large since $|\<\mu_{\beta_k}|\mu_{\beta_{k+1}}\>|^2 \ge \frac{1}{15}$.
  Therefore, we can use the method described in Fact~\ref{lem:jumpbymeasure} to ``jump'' from $|\mu_{\beta_k}\>$ to $|\mu_{\beta_{k+1}}\>$. The necessary projective measurements can be implemented with the reflections around the Gibbs qsamples. Since there are at most $\sqrt{q \ln n}$ stages, we want the failure probability of any jump to be smaller than $(\delta/2)/\sqrt{q \ln n}$ to ensure that the probability that any jump fails is $\le\delta/2$. The number of reflections across the Gibbs qsamples is 
  $O\big( \sqrt{q\ln n} (\ln q + \ln n)(\ln q + \ln \ln n + \ln(1/\delta)) \big)$. 
\end{proof}

Before we prove the main theorem in this section, we need the following result about the sample complexity of computing functions of random variables, given access to coherent qsamples of distributions. The proof of this theorem has been deferred to Appendix~\ref{app:nondestructiveproof}.

\begin{theorem}[Bounded relative variance]\label{thm:estimatingboundedvariance}
Suppose a distribution $D:\Omega\rightarrow [0,1]$ and function $f:\Omega\rightarrow [0,\infty)$ satisfy $B$-bounded relative variance for $B>1$.  
Then, there is a quantum algorithm $\A$ that: given $16 B \ln \big( 2/\eta\big)+1$ copies of~$\ket{\psi_D}$, with probability at least $1-\eta$, $\A$ outputs an $\varepsilon$-relative estimate $\hat{\mu}$ of $\mu=\mathbb{E}{x\sim D}[f(x)]$.
Additionally, $\A$ restores one copy of $|\psi_D\>$ and
invokes the reflection~$R_D$ (i.e., the reflection across the state $\sum_{x\in\Omega} \sqrt{D(x)} \ket{x}$)
$$
O\Big (\sqrt{B}/\varepsilon \cdot \big(\ln(B/\varepsilon) \big)^{1.5} \cdot
    \ln\big(\ln(B/\varepsilon) / \eta \big)  \Big)
$$
many times.
\end{theorem}

The theorem above allows us to estimate the expectations $\mathbb{E}[W_k]$ and $\mathbb{E}[V_k]$ by setting $D$ to be the Gibbs distribution $\mu_{\beta_k}$ and $f$ to be $\exp(-\bar{\beta}_{k,k+1} H(x_k))$ for the random variable $V_k$ or to $\exp(\bar{\beta}_{k,k+1} H(x_{k+1}))$ for the random variable $W_k$, as defined in Section~\ref{sec:pairproduct}. We can use that $B\le 15$ because the cooling schedule $\beta_0,\ldots,\beta_\ell$ generated by the algorithm in Theorem~\ref{thm:qschedulegen} is $15$-slowly varying.  The error parameter $\varepsilon$ and the failure probability $\eta$ will have to be of the order of $1/\ell$.

\subsection{Quantum algorithm}

\begin{theorem}
\label{thm:quantummainthm}
Let $n\geq 1$, $\delta,\varepsilon,\eta\in (0,1)$ and $0\leq \beta_{\min}<\beta_{\max}$. Let $H:\Omega\rightarrow [0,n]$ be a classical Hamiltonian. Let $Q=\Zp(\beta_{\max})/\Zp(\beta_{\min})$ and $q=\ln Q$. There exists a quantum algorithm that uses $\tilde{O}\left(q\cdot \varepsilon^{-1} \right)$
many applications of reflection operators around Gibbs state (at different inverse temperatures) and with probability at least $4/5$, approximates $Q$ up to to relative error $\varepsilon$, i.e., outputs $\widehat{Q}$ such that
$
    (1 - \varepsilon) \cdot Q\le \widehat{Q}\leq (1 + \varepsilon) \cdot Q
$
\end{theorem}
\begin{proof}
The proof of the quantum algorithm is very similar to the classical proof in Theorem~\ref{thm:classicalmainthm}. The main differences are in estimating the cooling schedule using quantum techniques and estimating the mean values using quantum techniques. Overall, the structure of the quantum algorithm is:
\begin{enumerate}
    \item Compute a slowly varying cooling schedule of length $\ell$ as described in Theorem~\ref{thm:qschedulegen}.
    \item Use the quantum mean estimation algorithm to estimate the expectations $\E[W_i]$ and $\E[V_i]$ with relative error $\varepsilon/\ell$ using Theorem~\ref{thm:estimatingboundedvariance}.  

    \item Multiply the estimates of $\E[W_i]$ to obtain an estimate of $\E[W]$ and  $\E[V_i]$ to obtain an estimate of $\E[V]$ and output their ratio as the final estimate.
    \end{enumerate}

We argue the correctness of the algorithm first. Step $(1)$ is clear. Theorem~\ref{thm:qschedulegen} satisfies the following: with probability $\geq 9/10$, the algorithm generates a sequence of $\ell=\sqrt{q\ln n}$ inverse temperatures $\beta_1,\ldots,\beta_\ell$ satisfying
\begin{align*}
    \varS[W_i]=\varS[V_i]&= \frac{\Zp(\beta_i)\Zp(\beta_{i+1})}{\Zp\left(\frac{\beta_i + \beta_{i+1}}{2} \right)^2} \leq 15 \qquad \text{ for every } i\in [\ell].
\end{align*}

Additionally observe that
$
\frac{\prod_i \mathbb{E}[V_i]}{\prod_i \mathbb{E}[W_i]}= \frac{\Zp(\beta_{\max})}{\Zp(\beta_{\min})} =Q.
$
One can use Theorem~\ref{thm:estimatingboundedvariance} which uses $O(\ln \ell)$ samples of each  $\ket{\mu_{\beta_i}}$ and invokes the reflection $R_{\beta_i}$ $\tilde{O}\big(\ell/ \varepsilon\big)$ many times and: with probability $\geq 1-1/(20\ell)$ produces a $\widehat{W}_i,\widetilde{V}_i$ such that 
\begin{align*}
    1-\varepsilon/(2\ell) \le \frac{\hat{V}_i}{\E[V_i]} \le 1+\varepsilon/(2\ell) \quad \mbox{and} \\
    1-\varepsilon/(2\ell) \le \frac{\hat{W}_i}{\E[W_i]} \le 1+\varepsilon/(2\ell). 
\end{align*}
We use ratios of the lower and upper bounds to bound the ratio $\hat{W}_i/\hat{V}_i$ from below and above and employ the union bound to obtain a bound on the success probability. We obtain that
\begin{align}
\label{eq:firstunionboundexp}
    (1-\varepsilon/(2\ell))^2 \le \frac{\hat{W}_i / \hat{V_i}}{\E[W_i] / \E[V_i]} \le 
    (1+\varepsilon/(2\ell))^2.
\end{align}
hold with probability at least $1-1/(10\ell)$ for all $i$. This in turn implies that
\begin{align}
\label{eq:finalvarianceestimate}
    (1-\varepsilon/(2\ell))^{2\ell} \le 
    \frac{\prod_i (\hat{W}_i / \hat{V_i})}{\prod_i (\E[W_i] / \E[V_i])} \le 
    (1+\varepsilon/(2\ell))^{2\ell}.
\end{align}
holds with probability at least $1-1/10-1/10\geq 4/5$ (the first $1/10$ is from the union bound over the $1-1/(10\ell)$ in satisfying Eq.~\eqref{eq:firstunionboundexp} and the second $1/10$ comes because Theorem~\ref{thm:qschedulegen} was assumed to fail with probability $\leq 1/10$). Recall that $\prod_i (\E[W_i] / \E[V_i]) = Q$, where $Q=\Zp(\beta_{\max}) / \Zp(\beta_{\min})$ is the desired ratio.
We obtain:
    \begin{align*}
    (1 - 2\varepsilon) \cdot Q\le
        (1-\varepsilon/(2\ell))^{2\ell}\cdot Q  \le 
    \prod_i (\hat{W}_i / \hat{V_i}) \le  \\
    (1+\varepsilon/(2\ell))^{2\ell} \cdot Q \leq 
    e^{\varepsilon} \cdot Q \le 
    (1 + 2\varepsilon) \cdot Q,        
    \end{align*}
    where the first inequality used $(1+x)^t\geq 1+xt$ (for $x\geq -1$ and $t\geq 2$), the  second and third inequality used Eq.~\eqref{eq:finalvarianceestimate}, the fourth inequality used $(1+x)^t\leq e^{xt}$ (for $x,t\geq 0$), last inequality used $e^x\leq 1+2x$ (for $x\in [0,1]$). Hence our final output $\prod_i (\hat{W}_i / \hat{V_i})$ is a $(2\varepsilon)$-relative estimator  of $Q$.

    It remains to analyze the complexity of this quantum algorithm. First, Theorem~\ref{thm:qschedulegen}~uses 
    $$
    O\Big(\ell \cdot (\ln q + \ln n) \cdot \ln \ell \Big)
    $$
    invocations of a reflection around the Gibbs state 
    (note that we fix $\delta=O(1)$ when invoking this~theorem).

    Next, Theorem~\ref{thm:estimatingboundedvariance} uses $O(\ell \ln \ell )$ copies of the Gibbs states and invokes the reflection operator around Gibbs state (at different inverse temperatures) $\tilde{O}\big(\ell^2/ \varepsilon\big)$ many times (we specify the poly-logarithmic factors below).  
     We can use Fact~\ref{lem:jumpbymeasure} and similar arguments as at the end of the proof of Theorem~\ref{thm:qschedulegen} to analyze the complexity of preparing these copies.  
    Observe that we cannot directly prepare a qsample $|\mu_{\beta_k}\>$ for an arbitrary $k$, but have to move successively through the sequence $|\mu_{\beta_0}\>, \ldots, |\mu_{\beta_k}\>$. In total, we need to make at most $O(\ell^2\ln \ell)$ transitions between Gibbs qsamples of adjacent stages by performing reflections around these qsamples. To use the union bound, we need that the failure probability of any transition should be on the order of $\ell^2\ln \ell$, which can be accomplished with at most $O(\ln \ell)$ many reflections per transition.     Therefore, $O\big( \ell^2\cdot  (\ln \ell)^2\big)$ many reflections are necessary in total to prepare all the copies.     Overall, using $\ell=\sqrt{q\ln n}$, our algorithm uses
    $$
    O(q+\ell^2\cdot  (\ln \ell)^2)=O(q\ln n\cdot (\ln q+\ln n)^2)
    $$
    walk steps to prepare these Gibbs state and
    \begin{align*}
O\Big(q/\varepsilon\cdot \ln n\cdot \big(\ln (q/\varepsilon \cdot \ln n)\big)^2\Big)
    \end{align*}
    invocations of the reflection around Gibbs state. The total number of quantum walk steps~is
    $$
    O\Big(q/\varepsilon\cdot \ln n\cdot \big(\ln (q/\varepsilon \cdot \ln n)\big)^2+q\ln n\cdot (\ln q+\ln n)^2\Big).
    $$
\end{proof}

    Finally, in order to translate the complexity stated in Theorem~\ref{thm:quantummainthm} into a running time bound, observe that each invocation of the reflection operator in the theorem involves $O(1/\sqrt{\Delta})$ Markov chain steps on top of the sample complexity above: this is a well-known result from~\cite{szegedy2004quantum,magniez11}. 

\begin{corollary}\label{thm:quantummain_Corr}
In the setting of Theorem~\ref{thm:quantummainthm}, there exists a quantum algorithm that makes $\tilde{O}\left(q\cdot \varepsilon^{-1}\cdot \Delta^{-1/2} \right)$ 
 steps of the quantum walk operator (each with spectral gap lower bounded by $\Delta$) and with probability at least $4/5$, approximates $Q$ up to to relative error $\varepsilon$, i.e., outputs $\widehat{Q}$ such that
$
    (1 - \varepsilon) \cdot Q\le \widehat{Q}\leq (1 + \varepsilon) \cdot Q.
$
\end{corollary}

\section{Conclusion}

First, we have improved a well-known classical algorithm for estimating partition functions. Second, we have shown how to quantize this improved algorithm, thereby obtaining a quadratic speed-up with respect to the estimation precision and to the spectral gaps of the underlying Markov chains.  In particular, we have obtained the best quantum algorithm for estimating partition functions, improving upon the state-of-the-art algorithm due to Harrow and Wei \cite{harrow2020adaptive}. 

\vspace{1mm}

\subsection{Open questions.} We conclude with a few concrete open questions. (1) Can we  \emph{fully} quantize the classical algorithms of~\cite{huber2015approximation} and Kolmogorov~\cite{Kolmogorov18}, in the process removing large prefactor in our algorithms? (2) Does there exist a classical algorithm for computing partition functions with schedule length $O(\sqrt{\ln |\Omega| \ln n})$, matching the conjecture by~\cite{vstefankovivc2009adaptive}? (3) Can we prove \emph{any} lower bound in the standard model of computation for computing partition functions (either quantum or classical)? \footnote{The lower bound of~\cite{Kolmogorov18} assumes a non-standard oracle access.} (4) Can one  remove the assumption that SVV or TPA applies only for non-negative Hamiltonians?

\section{Acknowledgment}

This work was partially supported by the IBM Research Frontiers Institute. SA, GN and KT acknowledge support from the Army Research Laboratory and the Army Research Office under grant number W911NF-20-1-0014. VH completed parts of this work at University of Bristol. VH received funding from the European Research Council (ERC) under the European Union’s Horizon 2020 research and innovation programme (grant agreement No.\ 817581) and the IBM PhD fellowship and wants to acknowledge discussions with Alex~Shestopaloff.

\bibliography{bibliography}
\bibliographystyle{plainurl}

\appendix

\section{Proof of schedule length}
\label{appendix:schedule_length}
We report the statement of Theorem~\ref{thm:promiseSVVschedule}, and detail its proof.
\label{app:schedulelength}
\begin{theorem}[Perfectly-balanced schedule length~\cite{vstefankovivc2009adaptive}]
There exists a sequence $\beta_0<\cdots<\beta_\ell$ with $\beta_0=\beta_{\min}$ and $\beta_\ell=\beta_{\max}$ satisfying the condition 
\begin{align}
    \label{eq:appaequality}
f\left( \frac{\beta_i + \beta_{i+1}}{2}\right) = \frac{f(\beta_i) + f(\beta_{i+1})}{2} - 1
\end{align}
and having length $\ell$ bounded from above by
\begin{equation*}
    \ell \le \sqrt{\big(f(\beta_{\min})-f(\beta_{\max})\big) \cdot
    \frac{1}{2} \ln\left( \frac{f'(\beta_{\min})}{f'(\beta_{\max})} \right)}.
\end{equation*}
\end{theorem}

\begin{proof}
Suppose we have already constructed the sequence up to $\beta_i$ and let $\beta_{i+1}$ be the largest value in $[\beta_i,\beta_{\max}]$ so that $\beta_i$ and $\beta_{i+1}$ satisfy Eq.~\eqref{eq:appaequality}. For notational simplicity, let
\begin{align*}
 \bar{\beta}_{i,i+1} = \frac{\beta_i + \beta_{i+1}}{2}, \quad   d_{i,i+1} = \frac{\beta_{i+1}-\beta_i}{2}, \quad  K_i      = f(\beta_i)-f(\beta_{i+1}).
\end{align*}
If $\beta_{i+1} \geq \beta_{\max}$, then we are done constructing the well-balanced sequence. Otherwise, by the maximality of $\beta_{i+1}$ with equality, we have
\begin{equation*}\label{eq:1}
    f(\bar{\beta}_{i,i+1}) = \frac{f(\beta_i) + f(\beta_{i+1})}{2} - 1.
\end{equation*}
Additionally since $f$ is convex,\footnote{In fact it is well-known that  the  partition function of a Hamiltonian is \emph{strongly convex}, not just convex (see~\cite{vuffray2016interaction}
for the proof of strong convexity for classical Hamiltonians). Using this strong convexity property it is possible that one could potentially improve the upper bound on the schedule length; we leave it as an interesting open question.} for every $a,b\in[\beta_{\min},\beta_{\max}]$ satisfying $a<b$, we have
\begin{equation}\label{eq:0}
    f'(a) \le \frac{f(b)-f(a)}{b-a} \le f'(b).
\end{equation}
 Setting $a=\beta_i$ and $b=\beta_{i+1}$ in Eq.\eqref{eq:0}, we have
\begin{equation*}
    f'(\beta_i) \le \frac{f(\beta_{i+1}) - f(\beta_i)}{\beta_{i+1}-\beta_i},
\end{equation*}
which we rewrite as
\begin{equation}\label{eq:2}
    -f'(\beta_i) \ge  \frac{K_i}{2d_{i,i+1}}.
\end{equation}
Setting $a=\bar{\beta}_{i,i+1}$ and $b=\beta_{i+1}$ in Eq.~\eqref{eq:0}, we obtain
\begin{equation*}
    \frac{f(\beta_{i+1}) - f(\bar{\beta}_{i,i+1})}{\beta_{i+1} - \bar{\beta}_{i,i+1}} \le f'(\beta_{i+1}).
\end{equation*}
Given $\beta_i,\beta_{i+1}$ satisfy Eq.~\eqref{eq:appaequality}, we have
\begin{align}
    -f'(\beta_{i+1}) 
    &\le 
    \frac{f(\bar{\beta}_{i,i+1}) - f(\beta_{i+1})}{d_{i,i+1}} =\frac{f(\beta_i) - f({\beta}_{i+1})-2}{2d_{i,i+1}}= \frac{K_i -2}{2d_{i,i+1}}. \label{eq:3}
\end{align}
Putting together Eq.~\eqref{eq:2},~\eqref{eq:3}, we have
\begin{equation}
    \frac{f'(\beta_{i+1})}{f'(\beta_i)} =
    \frac{-f'(\beta_{i+1})}{-f'(\beta_i)} \le
    \frac{K_i - 2}{K_i} = 1 - \frac{2}{K_i} \\
    \le \exp\left(-\frac{2}{K_i}\right).\label{eq:4}
\end{equation}
Taking the product of Eq.~\eqref{eq:4} for all $i\in\{0,\ldots,\ell-1\}$ 
and rearranging, we obtain
\begin{equation}\label{eq:5}
    \prod_{i=0}^{\ell-1} \exp\left(\frac{2}{K_i}\right) \le \prod_{i=0}^{\ell-1} \frac{f'(\beta_i)}{f'(\beta_{i+1})} =
    \frac{f'(\beta_{\min})}{f'(\beta_{\max})}.
\end{equation}
Taking the logarithm Eq.~\eqref{eq:5} and dividing both sides by $2$, we obtain
\begin{equation}\label{eq:6}
    \sum_{i=0}^{\ell-1} \frac{1}{K_i} = 
    \frac{1}{2} \ln\left(\frac{f'(\beta_{\min})}{f'(\beta_{\max})}\right). 
\end{equation}
Summing $K_i$ for all $i\in\{0,\ldots,\ell-1\}$, we obtain
\begin{equation}\label{eq:7} 
    \sum_{i=0}^{\ell-1} K_i = f(\beta_{\min}) - f(\beta_{\max}).
\end{equation}
Applying Cauchy-Schwarz inequality to Eq.~\eqref{eq:6},~\eqref{eq:7} we obtain
\begin{equation*}
    \ell^2 \le \big( f(\beta_{\min}) - f(\beta_{\max}) \big) \cdot 
    \frac{1}{2} \ln\left(\frac{f'(\beta_{\min})}{f'(\beta_{\max})}\right).
\end{equation*}
This concludes the proof of the theorem.
\end{proof}

\section{Bounding number of iterations in classical schedule generation}
\label{app:longmoveslemma}
Here we prove Lemma~\ref{lem:longmoves} (restated below for convenience) that bounds  number of ``long moves'' in  case 1  of the proof of Theorem~\ref{thm:findingSVVschedule}.
\begin{lemma}
The number of ``long moves'' in Algorithm~\ref{alg:cschedgen}, where we set $\beta^* = L^* = L$, is at most~$6 \sqrt{q} \ln n$.
\end{lemma}
\begin{proof}
We follow the same scheme as the proof of \cite[Lemma 5.14]{vstefankovivc2009adaptive}, with a few modifications to account for the differences between our Algorithm \ref{alg:cschedgen} and the schedule generation algorithm in \cite{vstefankovivc2009adaptive}. Because of these modifications, we are able to obtain a tighter upper~bound.

Recall that $I \subseteq \{0,\dots,n\}$ and $I$ is a contiguous interval, therefore the set of possible interval widths is a subset of $\{0,\dots,n\}$. We first consider long moves that happen with interval width of at least $1$; as discussed at the end of this proof, there can be at most one long move with interval width~$0$, therefore a bound on the number of long moves with interval width in $\{1,\dots,n\}$ immediately implies a bound on the total number of long moves. Let $x_k$ be the number of times that we perform a long move with an interval of width $k$, for $k=1,\dots,n$. Let $k_{\max} := \arg \max_k \{x_k : x_k > 0\}$, i.e., the maximum interval width $k$ for which we perform at least one long move.

Let us consider all the long moves performed with an interval width belonging to the set $\{k_{\min},\dots,k_{\max}\}$, where we arbitrarily fix a choice $k_{\min} \in \{1,\dots,k_{\max}\}$. We want to derive a lower bound on the inverse temperature $\beta$ reached after $x_{k_{\min}} + x_{k_{\min} + 1} + \dots + x_{k_{\max}} = t + 1$ such long moves.  Since $\beta$ is only increasing in the course of the algorithm, clearly a lower bound after $t$ moves is also valid after $t+1$ moves. In particular, since we want to obtain the tightest possible lower bound, we consider $\beta$ after performing $y_k$ moves with interval width $k$, where $y_k$ is defined as follows for $k \in \{k_{\min},\dots,k_{\max}\}$:
\begin{equation*}
    y_k = \begin{cases} x_k & \text{if } k < k_{\max} \\ x_k -1 & \text{if } k = k_{\max}. \end{cases}
\end{equation*}
Notice that this is equivalent to undercounting the number of moves at interval width $k_{\max}$ by exactly 1. Since for interval width $k$, the inverse temperature $\beta$ increases by exactly $1/k$ during a long move, after $t$ long moves that satisfy the move count $y_{k_{\min}},\dots,y_{k_{\max}}$ we have:
\begin{equation}
\label{eq:betalbint}
    \beta \ge \sum_{k=k_{\min}}^{k_{\max}} \frac{y_k}{k}.
\end{equation}
We claim that in addition, $\beta$ must satisfy:
\begin{equation}
\label{eq:betaubint}
    \beta \le \frac{q + \ln \frac{1}{h}}{k \sqrt{q}}.
\end{equation}
We now prove this claim. First, notice that by construction of $P$, since we choose the width $w$ of an interval $I = \{b,\dots,b+w = c\}$ to be $w = \floor{b/\sqrt{q}}$, we must have:
\begin{equation}
  \label{eq:intblb}
  b \ge w \sqrt{q}
\end{equation}
Next, notice that in order for a long move to happen starting from inverse temperature $\beta$ with an interval $I$ of width $w$, there must exist some inverse temperature $\beta' > \beta$ at which $I$ is $h$-heavy (more specifically, $I$ must be $h$-heavy at $\beta' = \beta + 1/w$, otherwise the long move cannot take place). The weight of $I$ at $\beta'$ is $\frac{1}{\Zp(\beta')} \sum_{x : H(x) \in I} e^{-\beta' H(x)}$. Therefore we must have:
\begin{equation*}
    h \le \frac{1}{\Zp(\beta')} \sum_{x : H(x) \in I} e^{-\beta' H(x)} \le \frac{1}{\Zp(\beta')} \sum_{x : H(x) \in I} e^{-\beta H(x)} \le \sum_{x : H(x) \in I} e^{-\beta H(x)},
\end{equation*}
where for the second inequality we used $\beta' > \beta$, and for the last inequality we used $\Zp(\beta') \ge 1$, which is true by assumption. We obtain the following chain of inequalities:
\begin{equation*}
    h \le \sum_{x : H(x) \in I} e^{-\beta H(x)} \le |\Omega| e^{-\beta k_{\min} \sqrt{q}},
\end{equation*}
where we used the facts that the width of $I$ is between $k_{\min}$ and $k_{\max}$, and Eq.~\ref{eq:intblb} (which, together, imply $H(x) \ge k_{\min} \sqrt{q}$ for the states considered in the summation). Taking the logarithm on both sides~yields
\begin{equation*}
    \ln h \le q - \beta k_{\min} \sqrt{q},
\end{equation*}
which immediately implies \eqref{eq:betaubint}. Combining \eqref{eq:betalbint} and \eqref{eq:betaubint}, we obtain:
\begin{equation}
\label{eq:yksumbase}
    \sum_{k=k_{\min}}^{k_{\max}} \frac{y_k}{k} \le \frac{q + \ln \frac{1}{h}}{k \sqrt{q}}.
\end{equation}
Now notice that $\sum_{k_{\min}=1}^{k_{\max}} \sum_{k=k_{\min}}^{k_{\max}} \frac{y_k}{k} = \sum_{k=1}^{k_{\max}} y_k$, because each term $\frac{y_k}{k}$ appears exactly $k$ times in the double summation. Therefore, taking \eqref{eq:yksumbase} and taking the sum for $k_{\min}=1,\dots,k_{\max}$ on both sides, we obtain:
\begin{equation}
\label{eq:yksumfinal}
    \sum_{k=1}^{k_{\max}} y_k = \sum_{k_{\min}=1}^{k_{\max}} \sum_{k=k_{\min}}^{k_{\max}} \frac{y_k}{k} \le  \sum_{k=1}^{k_{\max}} \frac{q + \ln \frac{1}{h}}{k \sqrt{q}} \le  (1 + \ln n) \frac{q + \ln \frac{1}{h}}{\sqrt{q}},
\end{equation}
where the last inequality exploits the fact that $k_{\max} \le n$ and the well-known inequality $\sum_{i=1}^n 1/i \le 1 + \ln n$. Using the choice $h = \frac{1}{8 |P|}$ in Algorithm \ref{alg:cschedgen} and Lemma~\ref{lem:partitionlength}, we can write $\ln \frac{1}{h} = \ln (32 \sqrt{q} \ln n) \le 2 \ln q \ln \ln n$. Finally, using \eqref{eq:yksumfinal} and $(1+\ln n) \le 2 \ln n$, we obtain:
\begin{equation*}
    \sum_{k=1}^{k_{\max}} y_k \le (2 \ln n) \frac{q + 2 \ln q \ln \ln n}{\sqrt{q}} \le 4 \sqrt{q} \ln n.
\end{equation*}
The LHS of the last equation is, by definition, equal to the number of long moves that are performed with interval width in $\{1,\dots,n\}$, minus one. Recall that in Algorithm \ref{alg:cschedgen} we can also have long moves with intervals of width $0$, but there can be at most one such move because then we set $L = q \ge \beta_{\max}$. Hence, the total number of long moves in the algorithm is at most $4 \sqrt{q} \ln n + 2 \le 6 \sqrt{q} \ln n$.
\end{proof}

\section{Proof of non-destructive amplitude estimation}
\label{app:nondestructiveproof}
We first introduce some definitions and notation which we use throughout. 

\begin{definition}\label{def:everything}
Let $\Omega$ be a finite set, $D: \Omega\rightarrow [0,1]$ be a  distribution, and $f : \Omega\rightarrow [0,\infty)$ be an arbitrary function. Define the mean 
$$
\mu=\Exp_{x\sim D} [f(x)]=\sum_{x\in\Omega} D(x) f(x)
$$ 
and the second moment 
$$
\phi=\Exp_{x\sim D} [f(x)^2]=\sum_{x\in\Omega} D(x) f(x)^2,
$$ 
and the relative variance $\frac{\phi}{\mu^2}$. We say \emph{$(D,f)$ satisfy $B$-bounded relative variance} if
$$
\frac{\phi}{\mu^2}\leq B.
$$
 We say that $\hat{\mu}$ is an \emph{$\varepsilon$-relative estimate} of $\mu$ if $|\mu-\hat{\mu}|\le \varepsilon \mu$.  
\end{definition}
Let $\ket{\psi_D}$ denote the \emph{coherent encoding} of the distribution $D$, i.e,
\begin{equation*}
    \ket{\psi_D} = \sum_{x\in\Omega} \sqrt{D(x)} \ket{x},
\end{equation*}
and $R_D$ the \emph{reflection} around $\ket{\psi_D}$, i.e.,
\begin{equation*}
    R_D = 2\ketbra{\psi_D}{\psi_D} - \id.
\end{equation*}

The results in this section are stated as general subroutines. Note that when we apply these techniques for the task of  estimating partition functions, the probability distribution $D$ corresponds to the Gibbs distribution $\mu_{\beta_i}$ at $\beta_i$, the function $f$ corresponds to the function $\exp(-\beta H(\cdot))$.\footnote{To be specific, $f$ will either correspond to $\exp(-\bar{\beta}_{i,i+1} H(x))$ for the random variable $W_i$ or to $\exp(\bar{\beta}_{i,i+1} H(x))$ for the random variable $V_i$, where these quantities are defined in Section~\ref{sec:pairproduct}. 
}

The goal of this section is to present a quantum algorithm for estimating $\mu$ with relative error~$\varepsilon$ given access to copies of the coherent encoding $|\psi_D\>$ and reflection operator $R_D$.  We assume that the relative variance $\frac{\phi}{\mu^2}$ is bounded from above by $B>1$. Naturally, the goal is to estimate $\mu$ using as few copies of $\ket{\psi_D}$ and invocations of the reflection operator $R_D$ as possible. Naively, a quantum algorithm could simply use multiple copies of $\ket{\psi_D}$ to obtain $O(B/\varepsilon^2)$ samples $x$ according to $D$, and estimate $\mu$ using the classical Chebyshev's inequality. However, quantumly one can obtain a quadratic improvement in $1/\varepsilon$, which we prove in this section. 

\begin{theorem}[Bounded relative variance]\label{thm:mainlemappendixB}
Suppose a distribution $D:\Omega\rightarrow [0,1]$ and function $f:\Omega\rightarrow [0,\infty)$ satisfy $B$-bounded relative variance.  
Then, there is a quantum algorithm $\A$ that: given $16 B \ln \big( 2/\eta\big)+1$ copies of~$\ket{\psi_D}$, with probability at least $1-\eta$, $\A$ outputs an $\varepsilon$-relative estimate $\hat{\mu}$ of $\mu=\Exp_{x\sim D}[f(x)]$. 
Additionally, with probability $\geq 1-\eta$,  $\A$ restores one copy of $\ket{\psi_D}$. Overall $\A$ 
invokes the reflection $R_D$ 
$$
O\Big(  \sqrt{B} / \varepsilon \cdot \big(\ln(B/\varepsilon) \big)^{1.5} \cdot
    \ln\big(\ln(B/\varepsilon) / \eta \big) \Big)
$$
many times.
\end{theorem}

The theorem above builds upon and improves the results due to \cite[Algorithm~4 and Theorem~6]{montanaro2015quantum}.  Most importantly, the complexity of our algorithm grows only with $\sqrt{B}/\varepsilon$, whereas the algorithm in \cite{montanaro2015quantum} grows with $B/\varepsilon$. 
Note that \cite{hamoudi2019chebyshev} already proved that a scaling with $\sqrt{B}/\varepsilon$ is possible and referred to this result as the quantum Chebyshev inequality.  Our approach provides a different algorithm, with a simpler proof, that achieves essentially the same running time. We also note that \cite{harrow2020adaptive} used the algorithm in \cite{montanaro2015quantum} as a subroutine for estimating partition functions.

 In order to prove this theorem, we will need a few theorems which we state first.  
 We first state  \emph{non-destructive} amplitude estimation: a  variant of amplitude estimation where the initial resource state is not destroyed in the quantum algorithm.\footnote{We remark that~\cite{chakrabarti2019quantum} also contains a procedure that performs non-destructive amplitude estimation.} 

\begin{theorem}[{Non-destructive amplitude estimation~\cite[Theorem~6]{harrow2020adaptive}}]
  \label{thm:ndampest_strong}
 Let $|\psi\>$ be an arbitrary quantum state and $P$ an arbitrary projector. Let $R_\psi=2 |\psi\>\<\psi| - I$. For every $t>0$, there is a quantum algorithm $\A$ that starts in the initial state $|\psi\>$ and with probability at least $1 - \eta$, outputs an estimate $\hat{p}$ of $p=\<\psi|P|\psi\>$ such that
\begin{equation*}
    |\hat{p} - p| \le 2\pi \frac{\sqrt{p(1-p)}}{t} + \frac{\pi^2}{t^2}.
\end{equation*}
Additionally, $\A$ restores $|\psi\>$ with probability $1-\eta$. Overall $\A$ invokes the controlled reflection  $R_\psi$ $O\big( t \cdot \ln (1/\eta) \big)$ many times.  
\end{theorem}

An immediate corollary of this theorem is the following.
\begin{corollary}
\label{cor:meanstimation}
Suppose a distribution $D:\Omega\rightarrow [0,1]$ and function $f:\Omega\rightarrow [0,1]$ satisfy $B$-bounded relative variance (for some $B\geq 1$). For every $t>0$, there is a quantum algorithm $\A$ that, given access to a single copy of $\ket{\psi_D}$ and controlled reflection $R_D$, with probability $\geq 1-\eta$ outputs an estimate $\hat{\mu}$ of $\mu$ such~that
\begin{equation}
    |\mu-\hat{\mu}| \le 2\pi \frac{\sqrt{\mu(1-\mu)}}{t} + \frac{\pi^2}{t^2}.
\end{equation}
Additionally, $\A$ restores the state $\ket{\psi_D}$ with probability $1-\eta$. Overall $\A$ invokes the controlled reflection $R_D$ $O(t \cdot \ln(1/\eta))$ many times. 

\end{corollary}
The algorithm in the corollary above is straightforward: first transform $|\psi_D\>$ into
\begin{equation*}
    |\psi_{D,f}\> = \sum_{x\in\Omega} \sqrt{D(x)} |x\> \otimes \big( \sqrt{f(x)} |1\> +  \sqrt{1-f(x)} |0\> \big).
\end{equation*}
by performing a controlled rotation on an additional qubit. 
For a  projector $P=\id\otimes |1\>\<1|$, observe that
$\<\psi_{D,f}|P|\psi_{D,f}\> = \sum_{x\in\Omega} D(x) f(x) = \mu$. So the algorithm can simply estimate $\mu$ using non-destructive amplitude estimation in Theorem~\ref{thm:ndampest_strong}.  The reflection around $|\psi_{D,f}\>$ can be realized with the help of the reflection $R_D$ and the controlled qubit rotation.

The lemma below provides an important subroutine required for proving Theorem~\ref{thm:mainlemappendixB}. The lemma makes use of the non-destructive amplitude estimation routine. Our proof follows closely \cite[Algorithm 2 and Lemma 4]{montanaro2015quantum}, but we choose a different way of partitioning $\Omega$, which allows us to reduce the complexity of the algorithm.

\begin{lemma}[Bounded second moment]\label{lem:boundedSecondMoment}
Suppose a distribution $D:\Omega\rightarrow [0,1]$ and function $f:\Omega\rightarrow [0,\infty]$ satisfy $B$-bounded relative variance. 
Then, there is a quantum algorithm $\A$ that, given  one copy of $\ket{\psi_D}$ and access to reflection operators $R_D$, with probability $\geq 1-\eta$ outputs~$\hat{\mu}$ such that
\begin{equation*}
|\mu-\hat{\mu}| \le \varepsilon.
\end{equation*}
Additionally, $\A$ restores the initial state $\ket{\psi_D}$ with probability $\geq 1-\eta$ and invokes the controlled reflection $R_D$  
$O\Big(  \sqrt{B}/\varepsilon  \cdot \big(\ln(B/\varepsilon) \big)^{1.5} \cdot \ln\big(\ln(B/\varepsilon) / \eta \big) \Big)$ 
 many times.
\end{lemma}

\begin{proof}
Let $k$ be a parameter to be determined later and consider the following sets
\begin{align*}
    \Omega_0     &= \{ x : 0          \le f(x) < 1 \}, \\
    \Omega_\ell  &= \{ x : 2^{\ell-1} \le f(x) < 2^\ell \} \quad \mbox{for} \quad \ell\in\{1,\ldots,k\}, \\
    \Omega_{k+1} &= \{ x: 2^k \le f(x) \}.
\end{align*}
We write the expectation
\begin{align}
    \mu 
    &=
    \sum_{\ell=0}^{k} 2^\ell \sum_{x\in\Omega_\ell} D(x) \frac{f(x)}{2^\ell} \,\,\, + \sum_{x\in\Omega_{k+1}} D(x)  f(x). \label{eq:mean}
\end{align}
The second term in~\eqref{eq:mean} can be bounded from above as follows
\begin{equation}
\label{eq:expxk+1}
    \sum_{x \in \Omega_{k+1}} D(x) f(x) \le 
    \frac{1}{2^k} \sum_{x\in\Omega} D(x) f(x)^2 
    \le \frac{B}{2^k},
\end{equation}
where we used $2^k \le f(x)$ for $x\in\Omega_{k+1}$.  We can ``ignore'' this term provided that $k$ is sufficiently large so its contribution to $\mu$ becomes negligible. 

Let us now focus on the double sum on the left in Eq.\eqref{eq:mean}, which can be understood as a weighted sum of $k+1$ means. For each $\ell\in\{0,\ldots,k\}$, we can estimate these means $\mu_\ell=\sum_{x\in\Omega_\ell} D(x) f(x) \cdot 2^{-\ell}$ with the help of Corollary~\ref{cor:meanstimation}.  To this end, we define the modified functions $f_\ell$ by setting $f_\ell(x)=f(x) \cdot 2^{-\ell}$ if $x\in\Omega_\ell$ and $0$ otherwise.  Let $\hat{\mu}_\ell$ be the estimates returned by amplitude estimation when applied to the probability distribution $p$ and the functions $f_\ell$. Our final estimate $\hat{\mu}$ of $\mu$ will then be
\begin{equation*}
    \hat{\mu} = \sum_{\ell=0}^k 2^\ell \cdot \hat{\mu}_\ell,
\end{equation*}

We now analyze the resulting estimation error. Let $t>0$ be a parameter determined later and let $\eta'=\eta/(k+1)$. We use Corollary~\ref{cor:meanstimation} (with parameters $t,\eta'$) to obtain $\hat{\mu}_\ell$ satisfying
\begin{equation}
\label{equ:promiseofamp}
    |\mu_\ell-\hat{\mu}_\ell| \le 2\pi \frac{\sqrt{\mu_\ell(1-\mu_\ell)}}{t} + \frac{\pi^2}{t^2}.
\end{equation}
Note that the algorithm in Corollary~\ref{cor:meanstimation} succeeds with probability $1-\eta'$, but since we invoke this corollary $k+1$ times for each $\ell\in \{0,\ldots,k\}$, by union bound the success probability is at least $ 1-\eta'\cdot (k+1)=1-\eta$. We have
\begin{align} 
    \sum_{\ell=0}^k 2^\ell |\mu_\ell - \hat{\mu}_\ell| 
    &\le
    \sum_{\ell=0}^k 2^\ell \left( \frac{2\pi \sqrt{\mu_\ell (1 - \mu_\ell)}}{t} + \frac{\pi^2}{t^2} \right) \notag\\
    &\le
    \sum_{\ell=1}^k 2^\ell \left( \frac{2\pi \sqrt{\mu_\ell}}{t} + \frac{\pi^2}{t^2} \right) +
    \frac{2\pi}{t} + \frac{\pi^2}{t^2} \notag\\
 &=\sum_{\ell=1}^k \left( \frac{2\pi \sqrt{ \sum_{x\in\Omega_\ell} D(x) 2^\ell f(x)} }{ t } + \frac{2^\ell\pi^2}{t^2} \right) + 
    \frac{2\pi}{t} + \frac{\pi^2}{t^2}
 \label{eq:usingdefnofmuell}\\
     &\le
    \frac{2\pi}{t} \sum_{\ell=1}^k \sqrt{ \sum_{x \in\Omega_\ell} D(x) f(x)^2 } + \frac{2\pi}{t} + \frac{\pi^2 2^{k+1} }{t^2} \label{eq:first}\\
    &\le
    \frac{2\pi}{t} \sqrt{k} \sqrt{\sum_{\ell=0}^k \sum_{x \in\Omega_\ell} D(x) f(x)^2 } + \frac{2\pi}{t} + \frac{\pi^2 2^{k+1}}{t^2} \label{eq:second} \\
    &\le
    \frac{2 \pi\sqrt{k} \sqrt{B}}{t} + \frac{2\pi}{t} + \frac{\pi^2 2^{k+1} }{t^2} \label{eq:hatmus}, 
\end{align}
where the first inequality follows from Eq.~\eqref{equ:promiseofamp},~\eqref{eq:usingdefnofmuell} used the definition of  $\mu_\ell=\sum_{x\in \Omega_\ell}D(x)f(x)\cdot 2^{-\ell}$,   Eq.~\eqref{eq:first} used $2^\ell\leq f(x)$ for $\ell\in\{1,\ldots,k\}$, Eq.~\eqref{eq:second} used Cauchy-Schwarz inequality and the final inequality used the relative variance upper bound (in the lemma statement). 

Using this, we can bound $|\mu-\hat{\mu}|$ as follows:
\begin{align}\label{eq:totalError}
    |\mu-\hat{\mu}| 
    &\le \sum_{\ell=0}^k 2^\ell |\mu_\ell - \hat{\mu}_\ell| + \sum_{x \in \Omega_{k+1}} D(x) f(x) \\
    &\leq 
    \frac{2 \pi \sqrt{k} \sqrt{B}}{t} + \frac{2\pi}{t} + \frac{\pi^2 2^{k+1} }{t^2} +  \frac{B}{2^k},
\end{align}
where we used the inequality in Eq.~\eqref{eq:hatmus} and the last inequality used Eq.~\eqref{eq:expxk+1}.
We now set the parameters $k$ and $t$ as follows. 

\begin{itemize}
\item Setting $k = \ln(2 B/\varepsilon)$ ensures that the fourth error term $B/2^k$ in Eq.~\eqref{eq:totalError} is at most $\varepsilon/2$. 

\item  Choosing $t\ge 4\pi\sqrt{B}/\varepsilon$ ensures that the third term $\pi^2 2^{k+1}/t^2$ in Eq.~\eqref{eq:totalError} is at most $\varepsilon/4$.

\item Finally, choosing $t\geq 8\pi (\sqrt{B} \sqrt{\ln(2B/\varepsilon)} + 1)/\varepsilon$ ensures that the sum $2\pi \sqrt{k}\sqrt{B}/t + 2\pi/t$ of the first two terms in Eq.~\eqref{eq:totalError} is at most $\varepsilon/4$.
\end{itemize}
We see that $k=O(\ln(B/\varepsilon))$ and 
$t=O(\sqrt{B}/\varepsilon\cdot \sqrt{\ln(B/\varepsilon}))$.
Overall, the number of times the reflection $R_D$ needs to be invoked is proportional to 
\begin{equation*}
    (k+1) \cdot t \cdot \ln\big( (k+1) / \eta \big) =
    O\Big ( \sqrt{B} / \varepsilon 
    \cdot 
    \big(\ln(B/\varepsilon) \big)^{1.5} 
    \cdot
    \ln\big(\ln(B/\varepsilon) / \eta \big) \Big).
\end{equation*}
This proves the lemma statement.
\end{proof}

We are now ready to prove the main theorem in this section.

\begin{proof}[Proof of Theorem~\ref{thm:mainlemappendixB}]
The goal is to estimate $\mu=\sum_{x\in\Omega} D(x) f(x)$. 
We first use \emph{classical} Chebyshev to obtain the following: take $16B$ qsamples $\ket{\psi_D}$, measure them in the computational basis to obtain $x\sim D$ and their mean produces a constant-factor estimate $\tilde{\mu}$ of $\mu$ such that
\begin{equation}\label{eq:roughEst}
    \mu/2 \le \tilde{\mu} \le 2 \mu
\end{equation}
with probability at least $3/4$.
In order to boost this probability: it is well-known that we can decrease the failure probability to $1-\eta/2$ simply by repeating the above process $\ln(2/\eta)$ many times and outputting the median of all the estimates.\footnote{For a proof of the powering lemma, we refer the interested reader to~\cite{jerrum86powering}.}  The total number of required samples used here is $m=16 B \ln \big( 2/\eta\big)$.

  However recall that the goal is to obtain an $\varepsilon$-approximation of $\mu$.  Suppose we have  a good estimate $\tilde{\mu}$ satisfying Eq.~\eqref{eq:roughEst}. Consider a rescaled function 
\begin{equation*}
    f_{\mathrm{res}}(x) = f(x)/\tilde{\mu}.
\end{equation*}
Its mean $\mu_{\mathrm{res}}$ and second moment $\phi_{\mathrm{res}}$ satisfy 
\begin{equation*}
    \mu_{\mathrm{res}} = \frac{\mu}{\tilde{\mu}} \quad \mbox{and} \quad
    \phi_{\mathrm{res}} = \frac{\phi}{\tilde{\mu}^2} \le \frac{4\phi}{\mu^2} \le 4 B,
\end{equation*}
where we used the lower bound on $\tilde{\mu}$ in Eq.~\eqref{eq:roughEst} and the assumption $\phi/\mu^2 \le B$ in the theorem statement. We now invoke the subroutine in Lemma~\ref{lem:boundedSecondMoment} for the distribution $D$ and  function $f_{\mathrm{res}}$ (clearly $(D,f_{\mathrm{res}})$ satisfies $B$-bounded variance): with probability $\geq 1-\eta/2$, this subroutine produces an estimate~$\hat{\mu}_{\mathrm{res}}$ satisfying
\begin{equation*}
    |\mu_\mathrm{res} - \hat{\mu}_{\mathrm{res}}| \le \varepsilon/2,
\end{equation*} 
which in particular also implies
\begin{equation*}    
 |\mu - \tilde{\mu} \cdot \hat{\mu}_{\mathrm{res}}|=    |\tilde{\mu}\cdot  \big({\mu}_{\mathrm{res}} -  \hat{\mu}_{\mathrm{res}}\big)|  \le \varepsilon/2\cdot  \tilde{\mu} \le \varepsilon \mu
\end{equation*}
so $\hat{\mu}=\hat{\mu}_{\mathrm{res}}$ is the desired $\varepsilon$-relative estimate of $\mu$.  By a union bound (over classical Chebyshev step and the subroutine in Lemma~\ref{lem:boundedSecondMoment}), the probability  of obtaining this $\varepsilon$-estimate is $\geq 1-\eta$. Overall the number of copies of $\ket{\psi_D}$ used is $16B\ln(2/\eta)+1$ (the first term is because of classical Chebyshev inequality and the second term is from Lemma~\ref{lem:boundedSecondMoment}). Moreover, note that since Lemma~\ref{lem:boundedSecondMoment} is non-destructive, one copy of $\ket{\psi_D}$ is restored in the process.  Additionally, the algorithm in Lemma~\ref{lem:boundedSecondMoment} uses the reflection operator $R_D$
$$
O\Big (\sqrt{B}/\varepsilon \cdot \big(\ln(B/\varepsilon) \big)^{1.5} \cdot
    \ln\big(\ln(B/\varepsilon) / \eta \big) \Big)
$$
many times.
\end{proof}

\end{document}